\documentstyle[aas2pp4,epsfig]{article}
\epsfverbosetrue

\def \sun {$_{\scriptscriptstyle \odot}$}
\def \ltaprx {\lower .1ex\hbox{\rlap{\raise .6ex\hbox{\hskip .3ex
        {\ifmmode{\scriptscriptstyle <}\else 
                {$\scriptscriptstyle <$}\fi}}}
        \kern -.4ex{\ifmmode{\scriptscriptstyle \sim}\else 
                {$\scriptscriptstyle\sim$}\fi}}}
\def\gtaprx {\lower .1ex\hbox{\rlap{\raise .6ex\hbox{\hskip .3ex
        {\ifmmode{\scriptscriptstyle >}\else 
                {$\scriptscriptstyle >$}\fi}}}
        \kern -.4ex{\ifmmode{\scriptscriptstyle \sim}\else 
                {$\scriptscriptstyle\sim$}\fi}}}

\newcommand{\sig}{\:\lower0.6ex\hbox{$\stackrel{\textstyle >}{\sim}$}\:}
\newcommand{\sil}{\:\lower0.6ex\hbox{$\stackrel{\textstyle
<}{\sim}$}\:}
\newcommand{\sigs}{\:\lower0.4ex\hbox{$\stackrel{\scriptstyle
       >}{\scriptstyle \sim}$}\,}
\newcommand{\sils}{\:\lower0.4ex\hbox{$\stackrel{\scriptstyle
       <}{\scriptstyle \sim}$}\,}

\begin{document}

\title{The Reaction Rate Sensitivity of Nucleosynthesis in Type II Supernovae}

\author{R. D. Hoffman$^{1,2,3}$ and S. E. Woosley$^{2,3,6}$}
\author{T. A. Weaver$^{2,3}$}
\and
\author{T. Rauscher$^{4}$ and F.-K. Thielemann$^{4,5}$}
\affil{$^1$Steward Observatory, University of Arizona, Tucson, AZ 85721}
\affil{$^2$Astronomy and Astrophysics Department, University of California,
Santa Cruz, CA 95064}
\affil{$^3$Lawrence Livermore National Laboratory, Livermore, CA 94550}
\affil{$^4$Department f\"ur Physik und Astronomie, Univ. Basel, CH-4056 Basel, Switzerland}
\affil{$^5$Institute for Theoretical Physics, University of California, Santa Barbara, CA 93106}
\affil{$^6$Max-Planck Institut f\"ur Astrophysik, 85740 Garching bei M\"unchen,Germany}

\begin{abstract}

We explore the sensitivity of the nucleosynthesis of intermediate mass
elements ($28 \le$ A $\ltaprx 80$) in supernovae derived from massive
stars to the nuclear reaction rates employed in the model. Two
standard sources of reaction rate data (Woosley et al. 1978; and
Thielemann et al. 1987) are employed in pairs of calculations that are
otherwise identical. Both include as a common backbone the experimental 
reactions rates of Caughlan \& Fowler (1988).
Two stellar models are calculated for each of two
masses: 15 and 25 M\sun. Each star is evolved from core hydrogen
burning to a presupernova state carrying an appropriately large
reaction network, then exploded using a piston near the edge of the
iron core as described by Woosley \& Weaver (1995). The final 
stellar yields from the models calculated with 
the two rate sets are compared and found to differ in most cases
by less than a factor of two over the entire range of nuclei
studied. Reasons for the major discrepancies 
are discussed in detail
along with the physics underlying the two reaction rate sets employed.
The nucleosynthesis results are relatively robust and less sensitive
than might be expected to uncertainties in nuclear reaction rates,
though they are sensitive to the stellar model employed.

\end{abstract}

\section{INTRODUCTION}

Two major efforts to model nucleosynthesis in massive stars have
operated in parallel for a number of years (e.g., Woosley \& Weaver
1995; Thielemann, Nomoto, \& Hashimoto 1996). These groups have differed in
the stellar physics they employed - convection theory, treatment of
stellar energy generation, starting point (the main sequence for the
former, helium cores for the latter), and how the supernova explosion
was simulated. They also differed, especially for elements heavier than
silicon, in their choice of nuclear reaction rates. Nomoto and his
collaborators have always used the Hauser-Feshbach calculations of
Thielemann et al. (1987; henceforth TAT).  So far, Woosley \& Weaver have
used those of Woosley et al. (1978; henceforth WFHZ).

Considering the many differences between the models, it is comforting
(perhaps even surprising) that the predicted nucleosynthetic yields
from these two groups agree as well as they do, but detailed
comparison does reveal discrepancies. Here we shall segregate
those differences that exist because of the dichotomy of stellar
models from those reflecting purely the choice of nuclear
physics. We shall also discuss the underlying assumptions in each
reaction rate set and why the two differ.  

We begin by describing ($\S$2) the two approaches to stellar evolution
and, briefly, the two rate sets (pending a more thorough review in
$\S$4). Section 3 presents new yields for some hybrid models - stars
evolved using the KEPLER code of Weaver, Zimmerman, \& Woosley (1978)
and exploded as described in Woosley \& Weaver (1995), but using rates
from both groups (TAT \& WFHZ). In addition to helping to understand
why calculations of the two groups differ, the use of two independent
sets of reaction rates in identical stellar models helps determine the
(nuclear physics portion of the) error bar one should assign to
nucleosynthesis studies of this sort. We find, in general that it is
relatively small.

In $\S$4, we examine in greater detail the nuclear physics that
produced the two reaction rate sets, specifically the input data and
physical assumptions employed in two globally parameterized
Hauser-Feshbach calculations by Woosley et al. (1978) and Thielemann
et al. (1987).  Previously unpublished comparisons to experimental
rates are given along with mutual comparisons of the two sets.

Section 5 then discusses the differences in stellar yields in light of
the results of $\S$4 as well as the known differences in the stellar
models. In $\S$6, we summarize and conclude.

\section{Previous Models and the Rates They Employ}

\subsection{A Tale of Two Stellar Models}
\label{tale2_sm}

The nucleosynthesis calculations provided to the community by
Thielemann, Nomoto \& Hashimoto (1996, hereafter TNH) are based on the
stellar evolution calculations of Nomoto \& Hashimoto (1988) for
massive stars. These calculations were extended to a larger grid of
progenitor masses by Hashimoto et al. (1993). Both works followed the
evolution of helium cores of fixed mass (omitting hydrogen burning and
complications of mass loss and red giant formation) and both used the
Schwarzschild criterion for convective instability.  The nuclear
energy generation and nucleosynthesis within a structural iterative
loop (stellar evolution time-step) was evaluated using a reduced
reaction network adopted to the appropriate burning stages (for
details see Thielemann, Hashimoto, \& Nomoto, 1990;
Nomoto \& Hashimoto 1988; Hashimoto, Hanawa, \& Sugimoto 1983;
and Hashimoto, 1995). 
A 30 isotope network was used from He to oxygen
burning that included essentially only stable nuclei up to Ca and did
not include neutron-rich nuclei.  Silicon burning, a much more
complicated stage of stellar burning in which equilibrated strong
flows must be followed and weak interactions on trace isotopes are
important, was treated in a different way. During the onset of
silicon burning a 148 isotope network was employed, extending to Ca
and including also unstable and neutron-rich species. After partial
burning of Si, when quasi-equilibrium in the Si-group was attained,
the calculation switched to using tables of the relevant quantities
evaluated previously with a 299 isotope network given in Thielemann,
Hashimoto, \& Nomoto (1990).  The tables gave for each instant of silicon
burning at density, $\rho$, temperature, $T$, and electron mole
number,$Y_e$, the appropriate energy generation rate, silicon
depletion rate, and time derivative of $Y_e$.  Technically this was
done by interpolating quantities from prior calculations for a grid of
densities, temperatures, $Y_e$, and Si mass fraction.  The individual
abundances of the isotopes could be obtained, if necessary, from
quasi-equilibrium at the same $\rho$, $T$, $X_{\rm Si}$, and $Y_e$.

Such a treatment led to an approximately correct energy generation
rate during all stages of stellar evolution and also kept track of the
major abundances. Limitations in this network, however, which e.g.
did not contain all production or depletion reactions of $^{26}$Al,
meant that it could not give reliable answers for the abundances of
such isotopes.  Similar limitations on the neutron-rich side of the
(stellar evolution) network did not permit the $s$-process to be
calculated.  ``Post-processing calculations" by Prantzos, Hashimoto \&
Nomoto (1990) gave an indication of what $s$-processing might be
expected during He-burning, but this study centered on the major
$s$-process nuclei beyond Fe (and not lighter elements like Si, P, S,
Cl, Ar, K, Ca, and Ti as discussed in K\"appeler et al. 1994)

This treatment differed in many ways from Woosley \& Weaver (1995, hereafter
WW95). These authors also made use of a smaller 19 isotope network to
calculate energy generation during hydrogen, helium, carbon, neon, and part
of oxygen burning. Though the matrix contained only 19 species, assumed
steady state links for about a dozen other nuclei made it the equivalent of a
larger network of about 30 isotopes (Weaver, Zimmerman, \& Woosley 1978).
After the oxygen abundance declined, during central oxygen burning, to below
4\%, the transition was made to a larger quasi-equilibrium network of 137
isotopes. This network was converged during each iterative cycle of the
stellar code and accounted for most of the computational expense of the
entire evolutionary calculation. In addition, Weaver et al. (and subsequently
WW95) used a different criterion for convective instability, the Ledoux
criterion plus semi-convection in the regions that were Schwarzschild
unstable but not Ledoux unstable. This contributed significantly to the
variable iron core masses obtained by the two groups for a given helium core
mass (in the Ledoux criterion a $Y_e$ barrier inhibits convection, in the
Schwarzschild criterion it does not). Finally, WW95 treated their large
network nucleosynthesis in a different way. A reaction network of 200
isotopes (large by the standards of the time) was carried in a
``co-processing" mode in each zone throughout the entire stellar evolution.
Thus the converged T and $\rho$ of each stellar model was used to update
this network at each time step, and most importantly, convective mixing was
carried out in the co-processed abundance set wherever appropriate. The
$s$-process up to A=65 was also calculated. This sort of detail was missing
in the presupernova models of Nomoto \& Hashimoto (1988) and Hashimoto et al.
(1993), and thus restricted the information available from the presupernova
evolution when it came time to calculate the explosion.

During the explosive part of the modeling, there were differences as
well, though not so dramatic. The size of the nuclear networks for this
phase have been discussed in WW95 and in
Thielemann, Hashimoto \& Nomoto (1990). Because of persistent
uncertainties in modeling the explosion of Type II supernovae as well as
the practical need to explore a wide range of stellar models, both groups
simulated explosions artificially. WW95 made use of
an inward (during in-fall, before the explosion) and outward (after the
induced explosion) moving piston with a velocity which caused a final
kinetic energy of the ejecta of $1.2 - 2 \times 10^{51}$~erg. The details of
this method were first described in Woosley \& Weaver (1986) and are
discussed in detail in WW95. Thielemann, Hashimoto
\& Nomoto (1990), Hashimoto et al. (1993), TNH,
and Nomoto et al. (1997) made use of a "thermal
bomb", i.e. introducing thermal energy in the form of a temperature
enhancement inside the Fe-core of the progenitor star. The thermal
energy was also chosen in such a way to attain a total kinetic energy
of the ejecta of $10^{51}$~erg.  While WW95
obtained a mass cut between ejecta and neutron star (which determined
the total amount of $^{56}$Ni ejected) from their choice of piston
position and energy (though this mass cut often lay well outside
their piston), TNH adjusted their mass cut based
upon existing observations which relate supernova progenitor masses to
the $^{56}$Ni ejecta. Both treatments have their advantages and
disadvantages (see Aufderheide, Baron, \& Thielemann 1991), 
neither is an adequate substitute
for a full understanding of the supernova mechanism. They are
parameterizations, but the parameters do reflect real physics, motion
inwards as well as outwards for the piston model, high entropy for the
innermost ejecta (reflecting the blowing of a bubble by neutrino
energy deposition) in the bomb model. 

\subsection{The Two Reaction Rate Libraries}
\label{t2_ratelibs}

In addition to the differences in stellar physics employed by both groups,
each employed an independent library of nuclear reaction rates.
These two data sets are a compilation of both experimental and
theoretical reaction rate information. The rate sensitivity problem
separates nicely into two mass regions, above and below silicon.

Below silicon, most of the important rates determining energy
generation and nucleosynthesis up to core carbon burning have been
measured directly in the laboratory
[Caughlan \& Fowler (1988, hereafter CF88); Caughlan et al. (1985); Harris et
al. (1983); and Fowler, Caughlan, \& Zimmerman (1975)]. 
Fits to these reaction rates have been adopted for some time by both
groups and form a common backbone of the two independent reaction rate
libraries with one important exception, the rate adopted for the
critical reaction $^{12}$C($\alpha,\gamma$)$^{16}$O.
WW95 used a fit to this reaction rate
recommended by CF88 but multiplied at all temperatures
by a constant factor of 1.7.
This corresponds to an $S$-factor(at 300 keV) of 170 keV barns,
a value which has been determined as optimal for producing the
solar abundance set (Weaver \& Woosley 1993; Timmes, Woosley \& Weaver 1995). 
Nomoto \& Hashimoto
(1988) chose the rate for $^{12}$C($\alpha,\gamma$) prescribed by
Caughlan et al. (1985), which corresponds at $T_9=0.3$ 
(i.e. core helium burning) to CF88 * 2.35.
Both of these values are compatible with the still existing
uncertainties of this rate (Buchmann 1997). An examination of the
impact on nucleosynthesis in the stellar models due to the particular
prescription for this critical rate is deferred to $\S$3.

In the heavier
mass range, reaction rates, except for the relatively well known
(n,$\gamma$) cross sections (Bao \& K\"appeler 1987; Wisshak et
al. 1997), are chiefly a product of Hauser-Feshbach theory.  For the
theoretical rates, there have basically been two choices available -
we present calculations using both, which will be loosely referred to
as ``TAT rates'' (Thielemann, Arnould, \& Truran 1987) and the ``WFHZ
rates'' (Holmes et al. 1976; Woosley et al. 1978). The corresponding
Hauser-Feshbach codes that calculated the nuclear cross sections 
from which these rate sets were constructed
are loosely referred to as ``SMOKER'' (TAT) and ``CRSEC'' (WFHZ, Holmes
1976).

The neutron capture reaction rates in both reaction rate libraries
incorporate the experimental cross section recommendations of
Bao \& K\"appeler (1987, for a more recent compilation see Beer, Voss, \& Winters,
1992). 
Above silicon, both groups have supplemented their Hauser-Feshbach 
reaction rates with results from experiment. The TAT set included
reactions on neutron- and proton-rich unstable targets
from Malaney \& Fowler (1988, 1989), Wiescher et al.
(1986, 1987, 1989), and Van Wormer et al. (1994).
Beyond this the TAT reaction rate library depended
strictly on rates derived from the Hauser-Feshbach calculations
of Thielemann, Arnould, and Truran (1987).

For charged-particle induced reactions on intermediate mass nuclei, the
experimental reaction rates implemented in the WFHZ reaction library
are mostly drawn from 
D.G. Sargood and his collaborators, Mitchell, Tingwell, Tims, Hansper,
and Scott. In most instances, this group has generated fits to 
reaction rates derived from their exhaustive cross section measurements,
these have been adopted in the WFHZ set wherever given.
When such fits were not available (Sargood 1982), the fits to the reaction
rates generated by the 
Hauser-Feshbach studies of Woosley et al. (1978) were normalized to agree
with experiment at temperatures appropriate to the nuclear burning 
stage where the
reaction rate was most important. For rates proceeding on targets heavier
than Si, this was usually $T_9=3.0$ (i.e. oxygen and silicon burning).
Comparisons between experiment and the Hauser-Feshbach results of 
both Thielemann et al. (1987) and Woosley et al. (1975, 1978) 
are presented in $\S$4.
A complete bibliography of all the
rates employed and the authors contributing to the WFHZ reaction rate
compilation used in this study, including an extensive tabulation of
these rates for $0.03 \le T_9 \le 10$, is given in Hoffman \&
Woosley (1992).  This data, plus the fit constants for the TAT rates,
are also available on the world wide web
(http://ie.lbl.gov/astro/astrorate.html).

For weak reactions, both groups always included as lower bounds to
the weak decay rates the {\sl ground-state rates} taken from the
measured laboratory values given in the {\sl Nuclear Wallet Cards}
(1990). Also included in all calculations were electron captures on
protons and positron captures on neutrons, which proved important. 
For temperatures above $5 \times 10^8$ K and densities
greater than 10$^5$ g cm$^{-3}$, all models used the weak interaction
rates of Fuller, Fowler, \& Newman (1980, 1982a,b, 1985) for all
nuclei lighter than mass 60. Above mass 60, no reliable tabulation of
the temperature and density dependence of the $\beta$-decay rates are
available, the ground state rates were used. 

Finally, neutrino induced reactions
on nuclei were included in the manner described in
WW95. These are based on the work of Wick Haxton
as described in  Woosley et al. (1990).
For select neutrino energies of 4.0, 6.0, \& 8.0 MeV,
tables of cross sections and particle
evaporation branching ratios for charged-current and neutral-current 
inelastic neutrino scattering reactions on intermediate mass nuclei 
are given by Hoffman \& Woosley (1992). 
Neutrino reaction rates have not been previously incorporated in either
TNH or Nomoto et al. (1997), but are included in all
models presented in this work.

\section{New Hybrid Results}
\label{hybrid_res}

To explore the dependence of the nucleosynthesis strictly upon the
choice of reaction rates above A=28, a set of new hybrid models was
generated that used the Woosley-Weaver choice of stellar physics, but
varied the reaction rate library.  Two stars having main sequence
masses of 15 and 25 M\sun \ were evolved from core hydrogen
burning to presupernova collapse and subsequent explosion.  Each
calculation was repeated twice, once with each rate set.  Since the
network employed for nucleosynthesis is decoupled from that used for
energy generation and $Y_e$ changes in the star, it was feasible to
split these operations. 

This also means that the energy generation network stayed the
same in both cases (including the nuclear input), which
guaranteed that the temperature and density conditions between two models
of the same mass but using different reaction rate sets stayed exactly the
same, thus only probing the rate uncertainties for the (co-processing)
nucleosynthesis. However, this did lead to a minor inconsistency 
between the rates
used in the small energy generation network used to converge the stellar
model and their counterparts in the large TAT reaction rate library utilized
for nucleosynthesis.
For all reactions on targets lighter than silicon, the
rates in both sets and in the energy generation network are
drawn from CF88, including the rate for 
$^{12}$C($\alpha,\gamma$)$^{16}$O (multiplied by 1.7) and the heavy ion-rates
($^{16}$O+$^{16}$O, etc.). 
For heavier species contained within the energy generation network, 
the Hauser-Feshbach rates of Woosley et al. (1978)
were used. This will have little effect on the results of the converged
stellar model, in that the structure of the star is mostly determined
prior to oxygen burning, where the CF88 rates will play the major role. 

\subsection{Stellar Yields for the WFHZ and TAT Reaction Rate Libraries}
\label{stel_ylds}

Figures \ref{fig1} and \ref{fig2} compare the detailed stellar
nucleosynthesis from four sets of calculations.
Plotted for each
stable isotope (after decay) is the ratio of the stellar yield (in
solar masses) from the models using the ``TAT rates'' divided by the
equivalent number obtained using ``WFHZ rates''.  
The stellar yields from the calculations using the WFHZ rate set 
should have been, and
in general are equivalent to those from models S15A and S25A of WW95.
The networks employed in each survey extended to $^{90}$Kr, (see Table
\ref{ntwks}), although the range of nuclei shown in the figures only
extends to $^{76}$Ge.  Isotopes of the same element are connected by solid
lines. The most abundant isotope of a given element is plotted as an
asterisk. Nuclei made chiefly as radioactive progenitors are
surrounded by diamonds, squares, or circles, depending on whether the
given nucleus was made by the same or different radioactive species as
a result of the different rate sets used in the model studied
(if a nucleus was chiefly produced as the same radioactive progenitor
by both reaction rate sets, the choice was a diamond).
The dotted lines indicate a factor of two difference in the ratio of the
stellar yield between the two models.  Overall the comparison is quite
good, much better than a factor of two in almost all cases. As we
shall see ($\S$4), the actual difference in reaction rates is
frequently larger than a factor of two, but there is compensation, in
that the major flows proceed through channels within the valley of
beta stability that are in good agreement between the two rate
compilations.

The nuclei which show greater than 20\% deviations in the 15 M$_\odot$
and 25 M$_\odot$ stars are: $^{33}$S, $^{40}$Ar, $^{40}$K,
$^{44,46}$Ca, $^{45}$Sc, $^{50}$V, and $^{70}$Zn. The causes for these 
discrepancies are analyzed is $\S$5.

\subsection{Comparison of Stellar Yields to TNH}
\label{comp_stel_ylds}

When considering a direct comparison of stellar yields between the two
groups (\cite{ww95}; \cite{tnh96}), apart from the many differences in
the separate treatments of stellar evolution ($\S$2), an important
distinction is the assumed rate for $^{12}$C($\alpha,\gamma$)$^{16}$O.
The rate for this reaction remains uncertain and yet plays a major
role in determining both the final structure of the presupernova star
and its ultimate nucleosynthesis (Weaver \& Woosley 1993).
TNH used the rate recommended by Caughlan et al. (1985), while
WW95 adopted the rate recommended by Caughlan \& Fowler (1998), 
but multiplied at
all temperatures by a constant factor of 1.7. In order to make a meaningful
comparison between these two models, as well as to explore the
effect of this critical rate, the 15 M\sun \ model of the previous section
was recalculated using both reaction rate sets, but assuming a
constant multiplicative factor for the CF88 
$^{12}$C($\alpha,\gamma$) rate of 2.35,
which agrees with the Caughlan et al. (1985) rate at T$_9=0.30$ during core
helium burning. This facilitates comparison between stellar models
that otherwise would have been very different.

The results are presented in Table \ref{new15vstnh96}.  For a solar
metallicity 15 M\sun \ star, three sets of stellar yields (in M\sun) are given,
those calculated with the WFHZ rate set, those with the TAT rate set, and
finally the yields from the 15 M\sun \ star of TNH. This list
covers the stable isotopes included in the TNH hydrostatic network up to oxygen,
and thereafter all stable isotopes (after decay) up to $^{70}$Ge. Also given
are two ratios of the stellar yields, which compare yields using the same
stellar model (WFHZ/TAT), and a common set of nuclear reaction rates
(TNH/TAT) in explosive burning. Blanks appear in place of entries that either
were not calculated by TNH or for which the comparison was less than 0.001.
The last column indicates the abbreviation for the 
dominant nucleosynthetic process(es)
responsible for production of the quoted stellar yield (or ratio of yields)
for a given isotope (see Table 19 WW95, also Woosley \& Weaver 1999).

Differences of more than 10\% are uncommon between
calculations that use a common stellar model (WFHZ/TAT), but larger differences
are apparent if the model varies (TNH/TAT). For the alpha-isotopes up to
$^{28}$Si the ratios are within a factor of three (typically better)
and are most likely due to the stellar model treatment (with the
different treatments of convection playing the dominant role).  Some agree
remarkably well ($^{20}$Ne, $^{26}$Mg, $^{31}$P, $^{44}$Ca, $^{48}$Ti, 
$^{60,61,62}$Ni), while others are very discrepant ($^{19}$F, 
$^{46,48}$Ca, $^{50}$Ti, $^{54}$Cr, $^{58}$Fe, and $^{64}$Ni).
These differences reflect either the physics inherent to each stellar model
or the limited reaction network used during stellar evolution phases in
TNH (discussed in $\S$2).

$^{14}$N and $^{23}$Na, for example, are products of hydrogen-burning,
which is not calculated in TNH, they started with helium cores.
The smaller differences within a factor of a few for nuclei
close to the stable N = Z line up to calcium are most likely due the stellar 
model treatment.  $^{19}$F (and to a lesser degree $^{26}$Al) are 
discrepant because TNH did not implement the $\nu$-process (Woosley et al 1990).
The limited network used by TNH during hydrostatic stellar evolution
likely accounts for the low production of many neutron-rich stable nuclei 
processed by
neutron-capture during the $s$-process ($^{36}$S, $^{40}$Ar, 
$^{46}$Ca, $^{58}$Fe). Yet other discrepancies
are due to their limited production in massive stars. 
$^{48}$Ca, $^{50}$Ti, and $^{54}$Cr, for example, 
are produced in nuclear statistical equilibrium in SN Type Ia  
(Woosley \& Eastman 1995). The yields quoted
for these species in the Kepler based models resulted from $s$-processing.

For nuclei predominantly made during explosive oxygen and silicon burning, the
agreement is often good ($\times 4$). An exception are most of the 
Cl-Sc isotopes, which typically differ by factors of 10 or more. 
These are susceptible to convective processing in the oxygen shell
shortly before collapse. The $s$-process can also contribute to species
in this mass range.

Table \ref{radkepvstnh} gives the important
gamma-line radioactivities made in each model;
$^{22}$Na, $^{26}$Al, $^{44}$Ti, $^{56}$Ni, $^{60}$Fe and $^{60}$Co.
(The quoted stellar yields have
been included in the appropriate stable isobars in Table \ref{new15vstnh96}). 
Production of $^{44}$Ti and $^{56}$Ni agree within a factor of two,
indicating that the different treatments affecting the mass-cut 
(moving piston vs. thermal bomb) achieve (in this case) consistent agreement 
between these two important supernova diagnostics.
The others disagree for reasons stated above. 
$^{60}$Fe and $^{60}$Co differ because they were made via
$s$-processing during neon-burning and in explosive processing in the 
Kepler based models. 
The $^{22}$Na agreement is good, but the yield is very small. 
We think this isotope is made predominantly in novae.
Convective processing in the neon-oxygen shell prior to collapse
(in the Kepler models) produced the larger $^{26}$Al yields.

We emphasize that the choice of reaction rates above silicon have less of
an effect on nucleosynthesis in massive stars than the stellar environment.
Improvements to both are clearly needed and welcome. 

\section{A Comparison between Reaction Rate Sets}

\subsection{Theoretical Nuclear Reaction Cross Sections}
\label{tnrcs}

We now move to a discussion of the {\sl nuclear} physics employed in
the two studies. The reaction rate sets employed by both contain, as
backbones, a number of experimentally determined values, the most
important being the charged-particle rates of Caughlan \& Fowler
(1988) and the neutron capture rates of Bao \& K\"appeler (1987). The
two sets differ however in the values adopted for the large number of
reactions that either cannot, or have not been measured in the
laboratory. A traditional theoretical approach in this mass range is
the statistical or Hauser-Feshbach model. This model is valid
only for high level densities in the compound nucleus, a condition
that is generally (but by no means universally) satisfied during the
advanced evolution of massive stars because of the heavy fuels and
high temperatures involved. The two compilations used respectively the
Hauser-Feshbach calculations of Holmes et al. (1976) and Woosley et
al. (1978) (WFHZ), or of Thielemann et al. (1987) (TAT). It is
therefore advisable to have a short look at the different approach of
these two nuclear models in order to understand the different results.

A high level density in the compound nucleus allows one to use
energy averaged
transmission coefficients $T$, which
describe absorption via an imaginary part in the (optical)
nucleon-nucleus potential (for details see Mahaux and
Weidenm\"uller 1979). This leads to the well known expression
\begin{eqnarray}
\label{statmod}
\sigma^{\mu \nu}_{i} (j,o;E_{ij})&=&
{{\pi \hbar^2 /(2 \mu_{ij} E_{ij})} \over {(2J^\mu_i+1)(2J_j+1)}} \nonumber \\
&\times & \sum_{J,\pi} {{(2J+1)}\over{T_{tot} (E,J,\pi)}} \nonumber \\
&\times & T^\mu_j (E,J,\pi ,E^\mu_i,J^\mu_i,\pi^\mu_i) \nonumber \\
&\times & T^\nu_o (E,J,\pi,E^\nu_m,J^\nu_m,\pi^\nu_m)
\end{eqnarray}
for the reaction $i^\mu (j,o) m^\nu$ from the target
state $i^{\mu}$ to the exited state $m^{\nu}$ of the final nucleus, with
center of mass energy E$_{ij}$ and reduced mass $\mu _{ij}$. $J$ denotes the
spin, $E$ the excitation energy, and $\pi$ the parity of excited states.
When these properties are used  without subscripts they describe the compound
nucleus, subscripts refer to the participating nuclei  $i$ and $m$ and
projectile and emitted particle $j$ and $o$ in the
reaction $i^\mu (j,o) m^\nu$, where the
superscripts indicate the specific excited states in the target nucleus $i$
and final nucleus $m$. 
Experiments measure $\sum_{\nu} \sigma_{i} ^{0\nu} (j,o;E_{ij})$,
summed over all excited states of
the final nucleus, with the target in the ground state. Target states $\mu$ in
an astrophysical plasma are thermally populated and the astrophysical cross
section $\sigma^*_{i}(j,o)$ is given by
\begin{eqnarray}
\label{stellar}
\sigma^*_{i} (j,o;E_{ij}) &=&
\sum_\nu \sigma^{\mu \nu}_{i}(j,o;E_{ij}) \nonumber \\
&\times & \sum_\mu (2J^\mu_i+1) \exp(-E^\mu_i /kT) \nonumber \\
&\times & {\left( \sum_\mu (2J^\mu_i+1)\exp(-E^\mu_i/kT) \right)}^{-1}
\end{eqnarray}
The summation over $\nu$ replaces $T_o^{\nu}(E,J,\pi)$ in Eq.(\ref{statmod}) by
the total transmission coefficient
\begin{eqnarray}
\label{trans}
T_o (E,J,\pi) &=
&\sum^{\nu_m}_{\nu =0}
T^\nu_o(E,J,\pi,E^\nu_m,J^\nu_m, \pi^\nu_m) \nonumber \\
&+& \int\limits_{E^{\nu_m}_m}^{E-S_{m,o}} \sum_{J_m,\pi_m}
T_o(E,J,\pi,E_m,J_m,\pi_m) \nonumber \\
&\times & \rho(E_m,J_m,\pi_m) dE_m.
\end{eqnarray}
Here $S_{m,o}$ is the channel separation energy, and the summation over excited 
states above the highest experimentally
known state $\nu_m$ is changed to an integration over the level density
$\rho$.
The summation over target states $\mu$ in Eq.(\ref{stellar}) has to be 
generalized
accordingly. 

In addition to the ingredients required for Eq.(\ref{statmod}), like the
transmission coefficients for particles and photons or the level densities, 
width fluctuation corrections $W(j,o,J,\pi)$ have to be
employed as well. They define the correlation factors with which all
partial channels of incoming particle $j$ and outgoing particle $o$,
passing through excited state $(E,J,\pi)$, have to be multiplied.
This is due to the fact that the decay of the state is not fully
statistical, some memory of the process of formation is retained and
influences the available decay choices. The major effect is elastic
scattering, the incoming particle can be immediately re-emitted before
the nucleus equilibrates. Once the particle is absorbed and not
re-emitted in the very first (pre-compound) step, equilibration is
very likely. This corresponds to enhancing the elastic channel by a
factor $W_j$. Besides elastic scattering, the main effect is a redistribution
of widths into weak channels, having a major impact at channel opening
energies when i.e. a (p,$\gamma$) and a (p,n) channel compete and the
weaker (p,$\gamma$) channel is enhanced.
Verbaatschot, Weidenm\"uller and Zirnbauer
(1984) derived a general expression 
in closed form, which is however computationally expensive 
to use. A fit to results from Monte Carlo calculations was given by
Tepel et al. (1974). While the width fluctuation
corrections of Tepel et al. (1974) are only an approximation to the
correct treatment, Thomas, Zirnbauer \& Langanke (1986) have shown that
they are quite adequate.
For a general discussion of approximation methods see Gadioli and
Hodgson (1992) and Ezhov and Plukjo (1993).
Of the presently available
thermonuclear rate sets in the literature (and used for nucleosynthesis
calculations) this effect is only included 
in the TAT rate set.  

The important ingredients of statistical model calculations
are the particle and $\gamma$-transmission coefficients $T$ and
the level density of excited states $\rho$. Therefore, the reliability 
of such calculations is determined by the accuracy with which these components 
can be evaluated. In the following we discuss
the methods utilized to estimate these quantities in the two sets of
reaction rates.

\subsubsection{Particle Transmission Coefficients}
\label{sec_trans}

The transition from an excited state in the compound nucleus $(E,J,\pi)$
to the state $(E^\mu_i,J^\mu_i,\pi^\mu_i)$ in nucleus $i$ via the emission of
a particle $j$ is given by a summation over all quantum mechanically allowed
partial waves
\begin{equation}
T^\mu_j (E,J,\pi,E^\mu_i,J^\mu_i,\pi^\mu_i) =
\sum_{l=\vert J-s \vert}^{J+s}
\sum_{s=\vert J^\mu_i -J_j \vert}^{J^\mu_i + J_j}
T_{j_{ls}} (E^\mu_{ij}).
\end{equation}
Here the angular momentum $\vec l$ and the channel spin $\vec s =\vec J_j+
\vec J^\mu_i$ couple to $\vec J = \vec l +\vec s$. The individual transmission
coefficients $T_l$ are calculated by solving the Schr\"odinger equation
with an optical potential for the particle-nucleus interaction. The 
studies of thermonuclear reaction rates by 
Holmes et al. (1976) and Woosley et al. (1978)
employed optical square well
potentials and made use of the black nucleus approximation. Thielemann
et al. (1987) used the optical potential for neutrons and
protons given by Jeukenne, Lejeune, \& Mahaux (1977), based on microscopic
calculations, with the local density
approximation. It included the corrections of the imaginary part by 
Fantoni, Friman, \& Pandharipande (1981)
and Mahaux (1982). 
For $\alpha$-particles they employed a
phenomenological Woods-Saxon potential
derived by Mann (1978), based on extensive data by McFadden \& Satchler
(1966). 

The differences between these treatments can be seen in Figures
\ref{s_trans_n}, \ref{s_trans_p}, and \ref{s_trans_a}, which show the
ratios of the TAT/WFHZ s-wave transmission coefficients 
for a set of stable nuclei with varying mass number A at a number
of particle energies (the large jumps in the ratios over a narrow range in A 
are due to this choice of abscissa,
with increasing mass one jumps from isotope to isotope as well
as element to element). The main differences are due to the black
nucleus approximation made by WFHZ.  This is equivalent to a fully
absorptive potential, once a particle has entered the potential well,
and therefore does not permit resonance effects.  If single particle
s-wave states are located close to the top of the potential well, they
can be experienced as resonances by incoming (s-wave) particles.  This
is shown for neutrons in Figure \ref{s_trans_n}. The maxima occur for
compound nuclei with 2$s_{1/2}$ and 3$s_{1/2}$ configurations at mass
numbers $A$$\approx$50 and 150.

Increasing particle energies leads to penetration of the potential 
well at energies further removed from the resonance energy for the same
target nuclei. Thus, the difference to the black nucleus approach, for the
nuclei experiencing the resonance behavior at low energies, vanishes and 
the ratio approaches unity. We see, however, that for thermal neutron energies 
of 0.1~MeV ($T\approx 10^9$~K), the difference is still quite remarkable
(dotted line, Figure \ref{s_trans_n}). 
Lighter nuclei will have the same (unpopulated) 2$s_{1/2}$ and 3$s_{1/2}$ 
configurations 
at higher excitation energies in the continuum, where they lead to very broad
resonances and therefore the ratio also approaches unity. 
This changes for p-wave 
neutrons, interacting with 1$p_{1/2,3/2}$ and 2$p_{1/2,3/2}$ configurations,
because the centrifugal barrier can still cause sharper resonances in the
continuum, and with increasing energy a shift of the peaks (which however
decline in size) to lower mass numbers is observed (not shown here).

Proton s-wave states below the Coulomb barrier, can, similar to the centrifugal
barrier behavior for p-wave neutrons discussed above, exist as resonances
for unbound particle energies (similarily the peaks shift with bombarding energy).
Therefore, strong resonance absorption still 
occurs for proton energies of 2-5 MeV (see Fig. \ref{s_trans_p}). 
In astrophysical applications these
are energies in the Gamow window, which correspond to temperatures of 
2-5~$\times 10^9$K.  

An additional effect, which is only pronounced for $\alpha$-particles,
can be seen in Fig.~\ref{s_trans_a}. The ratio of the transmission 
functions is systematically
rising with mass number. For lower $\alpha$-energies the rise starts at
lower mass number. This behavior can be understood in terms of
absorption in the Coulomb barrier (~\cite{mic70}). At sub-Coulomb energies
the particle wave function is increasing exponentially as a function
of radius within the barrier.
This has the effect that important contributions to the transmission
functions occur well beyond the normal nuclear radius at energies very
far below the Coulomb barrier (cf.\ Fig.\ 11 of \ \cite{mic70}). 
These contributions are not considered in
the treatment of square well potentials and the black nucleus
approximation, which assumes that absorption takes place inside the
nuclear radius. Therefore, the transmission is systematically
underestimated for sub-Coulomb energies.

It should be noted, however, that a correction factor $f$ is introduced
within the framework of square well potentials, which corrects for the
unphysical additional reflection at the edge of the potential well.
It is taken to be independent of energy which has been shown to be a
good approximation for nucleonic projectiles~(\cite{pea57,vog68}). 
However, it should be energy-dependent for $\alpha$-particles. It is an
interesting fact that the two approximations made in neglecting
absorption deep in the barrier and in employing an energy-independent
correction factor nearly cancel for energies not too much
below the barrier~(\cite{mic70}). 
That is why the ratio of the TAT and WFHZ $\alpha$-transmission
functions deviate from unity only for $A>60$, even for low
energies. (Historically, this is also the reason why an energy-dependent
correction factor for $\alpha$-particles was rarely used because it did
not improve the accuracy of the transmission functions.)

\subsubsection{$\gamma$-Transmission Coefficients}
\label{gam_trans}

The dominant $\gamma$-transitions (E1 and M1) have to be
included in the calculation of the total photon width.
The smaller, and therefore less important,
M1 transitions have usually been treated with the simple single particle
approach ($T \propto E^3$, Blatt and Weisskopf 1952), as discussed in
Holmes et al. (1976) and used in both compilations presented here. The 
E1 transitions are usually calculated on the basis of the Lorentzian 
representation of the Giant Dipole Resonance (GDR) [see the 
experimental compilations which are provided e.g. by  Berman
(1975), Berman and Fultz (1975), Carlos et al. (1974), 
Berman et al. (1979), and Gurevich et al. (1981)]. 
The E1 transmission coefficient
for the transition emitting a photon of energy $E_{\gamma}$ 
is a function of $E_{\gamma}$, the GDR energy $E_{G}$ and width $\Gamma_G$.
Holmes et al. (1976) and Woosley
et al. (1978) used analytical fits as a function of A and Z which are quite
accurate for $E_G$.
The (hydrodynamic) droplet model approach for $E_G$ by Myers et al.
(1977) gives an excellent fit to the GDR energies and
can also predict the split of the resonance for deformed nuclei, when
making use of the deformation. 

The width of the GDR is less understood even in modern 
Random Phase Approximation calculations
(Catara et al. 1997),
especially for the pronounced, observed shell effects with 
strongly reduced widths at magic numbers. Holmes et al.
(1976) and Woosley et al. (1978) used an analytical fit to simulate
this behavior.
Thielemann and Arnould (1983) proposed a phenomenological model for the GDR
width which satisfactorily reproduces the experimental data for spherical
and deformed nuclei and 
can be described as a superposition  of a macroscopic width due
to the viscosity of the nuclear fluid and a coupling to quadrupole surface
vibrations of the nucleus (see also Cowan, Thielemann \& Truran 1991).

Utilizing the methods outlined above for the total radiation
width at the neutron separation energy for all nuclei with experimental
information, Thielemann et al. (1987) found agreement generally within
a factor of 1.5. This was an improvement over earlier attempts
by Johnson (1977) and Hardy (1982). We expect from several tests that
the differences between the WFHZ and TAT gamma transmission functions is
not larger than a factor of 2.

\subsubsection{Level Densities}
\label{lev_dens}

Most statistical model calculations use the level density description of
the back-shifted Fermi gas (Gilbert and Cameron 1965)
\begin{equation}
\rho(U,J,\pi)={1 \over 2} f(U,J) \rho(U)\quad,
\end{equation}
with
\begin{eqnarray}
\rho(U) &=& {1 \over \sqrt{2\pi} \sigma}{\sqrt{\pi} \over
12a^{1/4}}{\exp(2\sqrt{aU}) \over U^{5/4}}\ \nonumber \\
f(U,J) &=& {2J+1 \over 2\sigma^2} \exp\left({-J(J+1) \over
2\sigma^2}\right) \nonumber \\
\sigma^2 &=& {\Theta_{\mathrm{rigid}}\over\hbar^2}\sqrt{U \over a}\ \nonumber \\
\Theta_{\mathrm{rigid}} &=& {2 \over 5}m_{\mathrm{u}}AR^2\ \nonumber \\
U&=&E-\delta
\end{eqnarray}
which assumes that positive and negative parities are evenly distributed
and that the spin dependence $f(U,J)$ is determined by the spin cut-off
parameter $\sigma$. The level density of a nucleus is therefore dependent
only on two parameters: the level density parameter $a$ and the back-shift
$\delta$. 
Gilbert and Cameron (1965) were the first to identify an empirical correlation
with experimental shell corrections $S(N,Z)$
\begin{equation}
\label{shcorr}
{a \over A} = c_0 + c_1 S(Z,N),
\end{equation}
where $S(N,Z)$ is negative near closed shells. 

Holmes et al. (1976) made an extensive tabulation of $a/A$ as a fit dependent
on N and Z in different nuclear mass ranges, which can be used to predict
the level density for any stable and unstable nucleus. Woosley et al. (1978)
also made use of a large amount of experimental level density information,
available for stable nuclei. This resulted in a set of experimental level
densities for such nuclei, which was utilized in their statistical model
calculations.

Thielemann et al. (1987) kept the same formalism as in Eq.(\ref{shcorr}) but
performed an independent evaluation of the coefficients $c_0$ and $c_1$.
By dividing the nuclei into three classes (a: those
within three
units of magic nucleon numbers, b: other spherical nuclei, c: deformed
nuclei), an improved agreement with respect to all previous approaches 
was obtained (maximum deviations of a factor
of 3 to 5 at the neutron separation energy).
This treatment was very phenomenological and unsatisfactory and still 
the weakest point in their procedures for the calculation of cross
sections.
Further improvements of level density predictions have been recently attained
(Rauscher, Thielemann \& Kratz 1997), but are not yet included 
in the rate
sets which are compared in the present paper.
In general deviations in level densities of up to a factor of 2-3 (if not 
more) have to be expected for the two rate sets. 

Reverse rates and rates of thermally populated targets include
partition functions of the participating nuclei. These
depend on the experimental knowledge of excited states.
As there is a time difference of about 10 years between the two compilations,
we expect the TAT set to be more up-to-date in this respect. Above the
highest know excited states, an integral over level densities has to be
performed. The above mentioned differences in the level density treatment 
would also apply to the partition functions.

\subsection{A Global Comparison}
\label{glob_comp}

Following the discussion in $\S$4.1.2 and 4.1.3,
we expect a
scattering of a factor 2-3 between the two rate sets due to the different
treatments of gamma widths and level densities.
On top of this, the systematic deviations due
to the different treatment of particle transmission coefficients, as discussed 
in section \ref{sec_trans}, will enter. It is educational to perform an
overall comparison. This is done in the following only for nuclei for which
the ground state properties are identical in both sets, and for reactions
that have identical Q-values. This limits the comparison to the vicinity
of the valley of beta-stability and permits a more clearly defined 
comparison. It should, however, be mentioned that the usage of different mass 
models for nuclei further from stability, entering reaction Q-values and ground
state properties, introduces additional uncertainties.

Hydrostatic oxygen burning is the first burning stage where a large number
of reactions with unstable targets enter nucleosynthesis calculations. These 
are rates which are not often contained in the experimental rate compilations.
The temperature for this burning stage is of the order $2\times 10^9$K.
The next burning stage, silicon burning, approaches 
already a nuclear statistical
equilibrium (NSE), where abundances are rather determined by chemical 
potentials (or binding energies or 
reaction Q-values) and not by the individual rates.
Temperatures around $2\times 10^9$K are also of importance for explosive 
burning. At (initially) higher temperatures in explosive environments, full 
NSE is usually attained and abundances depend again on masses or Q-values.
In an expanding and cooling medium
such equilibria freeze-out typically at temperatures of $\approx$$2.5\times
10^9$K. It is here where individual reaction rates count and have the
strongest influence. 
Thus, for various reasons, the temperature range around $2.5\times 10^9$K
is the most important for thermonuclear reaction rates of intermediate and
heavy nuclei, and
we chose this temperature for a general comparison.

\subsubsection{Comparison: Theoretical Rates}
\label{comp_theo_rates}
 
In order to understand the differences, it is instructive to
take a look at global trends in the comparison of the TAT and WFHZ
rates in Figures~\ref{ng}--\ref{ap}. The ratios of the TAT/WFHZ
rates are plotted for a large number of nuclei for
(n,$\gamma$), (p,n), ($\alpha$,n), and ($\alpha$,p) reactions.
Let us start with the discussion of (n,$\gamma$) rates. A temperature of
2.5$\times 10^9$K corresponds to thermal neutron energies of the order 0.25~MeV.
For such energies Figure \ref{s_trans_n} 
already indicates only small deviations of less than a 
factor of two for neutron transmission coefficients (somehwere between
the dotted and short-dashed line).
This is smaller than the remaining scatter in the comparison of the
(n,$\gamma$) rates, due to the different treatment of other 
properties, like gamma transmission coefficients and nuclear level densities.
Thus, we find a relatively flat deviation for the ratios of transmission
coefficients with an overlying scatter of a factor 2-3, as expected from
sections \ref{gam_trans} and \ref{lev_dens}. 
This scatter is not specifically pronounced as a function
of nuclear structure and close to a random behavior.

To avoid this general scatter, we will focus in the following on particle
channels, as they result from the use of different optical potentials,
where we expect more systematic differences.
For this we choose as a next example in Figure \ref{pn} (p,n) reaction rates.
A temperature of 2.5$\times 10^9$K corresponds in the Gamow window to an
energy of $\approx$2.5~MeV protons. For such energies we see in 
Figure \ref{s_trans_p} strong deviations from unity and expect a maximum around
$A$$\approx$70-80. This is exactly what can be seen in Figure \ref{pn}. 
We also see
a minimum around $A$$\approx$40-50 and again a rise to smaller mass numbers,
as expected from Figure \ref{s_trans_p}.

Given the quite different values for, e.g., the
proton transmission function (Fig. \ref{s_trans_p}), 
it is perhaps surprising that
the rates (e.g., for (p,n) reactions, Fig. \ref{pn}) differ so little.
This is because the (p,n) rates on heavy nuclei are
only astrophysically important at temperatures above about $3 \times
10^9$ K where the Gamow energy is above 2 MeV. Figure \ref{s_trans_p}
is only for s-waves, but other partial waves do contribute.
The difference in the proton transmission function ratio will decrease with
increasing partial wave because the angular momentum barrier will
dominate and this barrier is the same in both approaches. At lower
energies we expect that the rates would differ more because of two
reasons: The higher partial waves will be more suppressed and the
difference in s-wave proton transmission function ratio
becomes slightly more pronounced. However, when
going from 2 MeV to 1 MeV we would not expect much difference. Only at
even lower energies would an increase be noticeable but these are
strongly hindered by the Coulomb barrier.

In Figure \ref{an} we 
display the rate comparison for ($\alpha$,n) reactions. Again,
as the neutron transmission coefficient ratio behaves already flat for these
temperatures, we expect essentially to see the behavior of alpha transmission
coefficients. This is also exactly what can be observed 
(see the final paragraphs of \ref{sec_trans}).
The ratio
rises as a function of A, as also seen in Figure \ref{s_trans_a}, 
with the deviations from unity starting around $A$=60.
Figure \ref{ap} for the ($\alpha$,p) rates comparison combines the effects of
Figure \ref{pn} with a maximum at $A$=70-80 and Figure \ref{an} 
with a continuous rise
for $A$$>$60. We see deviations in this upper mass range which can amount 
to a factor 8. Up to $A\approx 55$ we still see a smaller scatter of a
factor 2-3.

Although the above effects of utilizing different potentials can already
be found for reactions with the target being in the ground state, they
may have further consequences in the calculation of stellar rates
involving thermally excited targets. As each thermally populated
target state gives rise to similar deviations as described above, the
temperature dependence of the rates will also be altered when comparing
WFHZ and TAT rates.

One additional difference should be mentioned here. TAT and WFHZ made
different assumptions about isospin mixing, which is most important
for Z=N nuclei, and accounts for the well known inhibition of dipole
transitions in self-conjugate (zero isospin) nuclei.  From the sparse
experimental data available at the time (Toevs, 1971; Rogers et
al. 1977; Dixon \& Storey, 1977; Cooperman, Shapiro, \& Winkler,
1977), Fowler and Woosley applied an empirical correction factor of
0.2 to the photon transmission function for these reactions. A less
dramatic suppression of 0.5 was assumed for (n,$\gamma$) and
(p,$\gamma$) reactions into self-conjugate nuclei, and a factor of
0.67 for all photon transmission functions in $|Z-N|=1$ nuclei.  There
is less experimental justification for the latter. These correction
factors were not included in the TAT rates nor
in the TNH nucleosynthesis calculations.  This accounts for the
roughly 30\% difference in the yield ratio for $^{44}$Ti ($\S2$). We
defer discussion of these issues as they relate to $^{44}$Ti synthesis
until $\S5$.
	
\subsubsection{Comparison: Theory vs. Experiment}
\label{comp_exp_rates}

An important test of any theoretical result is how well it compares to
experiment. The Hauser-Feshbach codes themselves are constructed with a
wealth of experimental data (isotopic binding energies,
ground state and excited state level energies and their associated
spins \& parities, level densities, etc.).
How this translates into reliable reaction rates
can only be determined by direct comparison to published experimental data.
In this section we compare cross sections 
predicted by the two Hauser-Feshbach
codes (``CRSEC'' for WFHZ \& ``SMOKER'' for TAT) and reaction rates
calculated from those cross sections, against their experimentally
measured counterparts.
The body of experimental reaction rate data was drawn from the
literature up to the year 1992 (for a complete list of authors and
reaction channels studied, see Hoffman and Woosley (1992) or the www address
previously listed).

Wherever experimental rate information is available, it should be used
in preference to a theoretical prediction. This was the principle
motivation of both the TAT and WFHZ groups in adopting the results
from the numerous reaction rate compilations
of W. A. Fowler and his collaborators for reactions on light and
intermediate mass nuclei ($A \ltaprx 30$).
Beyond this the number of stable species (for each element),
the large number of particle channels, and the important
reactions that proceed through them, coupled with the
experimental difficulties encountered when dealing with
Coulomb barrier penetration on progressively
heavier targets, conspire to
limit the number of experiments that have been carried out on targets heavier
than silicon.

We begin by considering measured neutron capture
cross sections. This will provide a valuable check on the reliability of
the photon-transmission functions adopted by the TAT and WFHZ efforts.
Shown in Figures \ref{ngcxl} and \ref{ngcxh}
are the ratios of the theoretical neutron capture cross sections (in mb
and evaluated at 30 keV) divided by their recommended experimental values
(Bao \& K\"appeler 1987) vs. mass number A=Z+N.
Where available, the error in the measured cross section
divided by the cross section itself
is indicated above and below unity (the dashed line) by error bars.
Cross section ratios (WFHZ/EXP) are given by
a filled diamond, (TAT/EXP) are given by an open square.
Pairs of ratios reflecting reactions
on a common target are coupled by a solid vertical line, the element of
the target is listed above the largest ratio for a given comparison, with
the mass number given on the abscissa.
Horizontal dotted lines are a factor of two above and below unity.
Not shown in Figure \ref{ngcxl} are the cross section ratios for two targets,
$^{26}$Mg (WFHZ/EXP=16.7, TAT/EXP=17.6), and 
$^{31}$P (WFHZ/EXP=4.5, TAT/EXP=4.7).
For $^{21}$Ne the theoretical to experimental cross section ratio 
for WFHZ is 1.8, while for TAT it is 2.8 (off scale).

We can calculate a statistical measure of the ability of the
Hauser-Feshbach codes to predict cross sections (and reaction rates)
over the range of available experimental data by determining the mean
and standard deviation of the value of the ratios for cross sections
predicted by theory divided by those measured by experiment.  We
restrict our comparisons to cross sections and reaction rates on
targets with $A\ge 28$. In general the level density for reactions on
lighter targets is too low for Hauser-Feshbach studies to be valid.
Moreover, both groups have used experimental reaction rates for
$A < 28$, we are testing our ability to model reactions on heavier
targets.

For the 30 keV cross sections (starting at $^{28}$Si) shown in 
Figures \ref{ngcxl} and \ref{ngcxh}
the mean (and standard deviation) of the cross section ratios 
predicted by SMOKER and CRSEC are 1.08 (0.64) and 0.91 (0.58) respectively.  
For the cross sections that are compared, both Hauser-Feshbach
codes predict (n,$\gamma$) cross sections that are close to
experiment nearly equally often. On average SMOKER predicts a cross section
that is larger than experiment, CRSEC predicts cross sections that are lower
than experiment.
The spread around the means is smaller for CRSEC, but not by much.
The somewhat smaller deviations for CRSEC in this
comparison with known experimental rates for stable nuclei also reflect to
some extent the utilization of experimental level densities
where available (see \ref{lev_dens}), i.e. predominantly for such nuclei where
experimental cross sections exist. Therefore, in predictions for unstable
nuclei where such information is not available, one should probably
expect (at best) accuracies similar to those resulting from SMOKER 
(where only theoretical predictions where used).

For charged-particle induced reactions on intermediate mass nuclei, the
experimental reaction rates implemented in the WFHZ reaction library
are mostly drawn from the efforts of
D.G. Sargood and his collaborators (see $\S2$ and Hoffman \& Woosley 1992).
We present comparisons for 63 experimentally measured charged-particle
induced reaction rates, separated into five reaction channels:
(p,n), (p,$\gamma$), ($\alpha$,p), ($\alpha$,n), and ($\alpha,\gamma$).
All comparisons will be made at one temperature, T$_9=3.0$.

Figures \ref{pncx} and \ref{pgcx} show ratios of
(p,n) and (p,$\gamma$) reaction rates that were 
calculated from the two 
theoretical cross section codes (SMOKER and CRSEC) 
divided by experimentally measured reaction rates
for protons incident on the targets listed by the element symbol in the
figure and the mass number on the abscissa. In Figure \ref{pncx}
the lightest reaction
compared is for $^{27}$Al(p,n)$^{27}$Si.
These figures are similar in all respects to Figures \ref{ngcxl} \&
\ref{ngcxh}, but
no error bars are given. Figures \ref{apcx} and \ref{ancx} show the
same ratios for alpha-induced reactions.
Statistical measures of the average error for Figures \ref{ngcxl} - \ref{ancx}
are compiled in Table \ref{thvexp}, which lists the number of theoretical
reaction rates
(or cross sections) compared to experiment, and
the mean and standard deviations of these ratios for each reaction channel 
studied.

For all reaction channels compared in Figures \ref{ngcxl} - \ref{ancx},
each theoretical code predicted
cross sections leading to reaction rates that agreed with experiment
roughly an equal number of times,
although for the (p,n) rates the mean error for the TAT/EXP reaction rate ratio is 
closer to unity and the spread about the mean is less.
For both codes, the error statistics are weighted heavily by the
two (p,n) rates on $^{45}$Sc and $^{50}$Ti, all other rates falling within
the ``factor of two" reliability lines.
If these ratios are removed from the
error analysis, the means (and standard deviations) 
for WFHZ/EXP and TAT/EXP reaction rate ratios
then become 1.24 (0.32) and 0.85 (0.40) respectively, with the mean again closer to unity
for the more recently calculated rates, although the spread around each mean is now
marginally better for the WFHZ set. This may reflect
the inclusion of more and better known excited state data, and superior
particle transmission functions inherent in the TAT treatment.

For the (p,$\gamma$) rates, the statistical errors indicate that CRSEC
predicted reaction rates closer to unity, but again the results
are weighted heavily by one problematic reaction rate.
$^{39}$K(p,$\gamma$)$^{40}$Ca has a WFHZ reaction
rate to experiment ratio of 3.22, while the TAT ratio is much larger, 9.23 (off scale 
in Fig. \ref{pgcx}). Neither Hauser-Feshbach code predicted
this rate with much accuracy, and if it is excluded from the error analysis the
statistics improve considerably, with the mean (and standard deviation) for
SMOKER and CRSEC being 1.33 (0.65) and 1.35 (0.81) respectively. 
These values are reported in Table \ref{thvexp}.
Another reaction (not considered in these comparisons) is  
$^{41}$K(p,$\alpha$)$^{38}$Ar,
which is endoergic (has a positive ``Q'' value in the reverse reaction
channel sense, i.e. decreasing charge or mass).
This (p,$\alpha$)-reaction rate (evaluated at T$_9=3.0$), as predicted by both the
SMOKER and CRSEC statistical codes, was underestimated by a
factor of three. Both of these cases are examples of possible deviations 
for reactions close to
shell closures where level densities and gamma transmission coefficients
undergo large changes.

For the alpha-induced reaction channels,
Figures \ref{apcx} and \ref{ancx} show that nearly all of the comparisons
fall within the "factor of two" reliability lines. 
The statistical errors for the ($\alpha$,n) rate comparisons from each code
are very
nearly the same, while for the ($\alpha$,p) comparisons they are only
marginally different (Table \ref{thvexp}).
Finally, for the ($\alpha,\gamma$) channel, we compare only two
experimentally measured reaction rates. For
$^{34}$S($\alpha,\gamma$)$^{38}$Ar, the reaction rate ratios
are WFHZ/EXP=1.35 and TAT/EXP=1.07.
For $^{42}$Ca($\alpha,\gamma$)$^{46}$Ti, the same numbers are 0.84 and
1.43 respectively. Both codes predict these rates well within a
factor of two uncertainty. 

It should be noted that for all reaction channels presented in Figures
\ref{ngcxl} through \ref{ancx} and Table \ref{thvexp}, the largest
discrepancies are often confined to a few rates, particularly
those either on (and especially leading into) 
targets with closed neutron shells. This is
where the two different treatments of the nuclear level density display the
largest uncertainties. Nuclear shell corrections are very pronounced
at or near magic nucleon numbers. Since reliable experimental data on
radiation widths are lacking in these regions, it makes it difficult to
disentangle the level density effects 
from a possible problem in the prediction of the gamma transmission
functions [although see also
the discussion concerning the assumed nuclear potentials 
(Equivalent Square Well vs. Woods-Saxon)
in $\S$ \ref{comp_stel_ylds} and Figures \ref{s_trans_n}-\ref{s_trans_a}].

Reactions on targets in the vicinity of N=20 have received special
attention by Sargood and his collaborators.
Specific examples are the proton-induced reactions on $^{50}$Ti, and
$^{52}$Cr. The $^{41}$K(p,$\alpha$)$^{38}$Ar rate leads into this same
closed shell. For the ($\alpha$,p) channel reactions on both
$^{48}$Ti and $^{50}$Ti, the statistical codes perform poorly.
Two exceptions are $^{34}$S($\alpha,\gamma$)$^{38}$Ar and
$^{54}$Fe($\alpha$,n)$^{57}$Ni, both were
well calculated by the statistical model codes.
It should again be noted here that all of the experimental reaction rate
information discussed above supplements the WFHZ reaction rate set 
which was utilized for the stellar yield predictions of $\S2$,
while TNH always made use of their theoretical 
TAT rates for intermediate mass nuclei.

\subsection{Future Improvements}
\label{fut_imp}

Based on the well-known statistical model code SMOKER
(Thielemann, Arnould, \& Truran 1987), an improved code for
the prediction of astrophysical cross sections and reaction rates in the
statistical model is presently developed.
In the new code NON-SMOKER (Rauscher and Thielemann 1998), transmission coefficients
for neutrons, protons and $\gamma$ transitions
are determined in the same way as
previously described in Section~\ref{sec_trans}.
Optical potentials for $\alpha$ particles are treated in the folding
approach of Satchler \& Love (1979), 
with a parameterized mass- and energy-dependence
of the real volume integral (Rauscher 1998). The mass- and
energy-dependence of the imaginary potential is parameterized according
to Brown and Rho (1981) and additionally includes microscopic and deformation
information (Rauscher 1998).

The level density treatment has been recently improved 
(Rauscher, Thielemann, \& Kratz 1997).
However, the problem of the parity distribution at low energies remains.
The new code includes this modified description, but also accounts for
non-evenly distributed parities at low energies, based on the most recent
findings within the framework of the shell model Monte Carlo method
(Nakada \& Alhassid 1997).

Additionally, the included data set of experimental level information
(excitation energies, spins, parities) has been updated
with the aid of the present version of the Evaluated
Nuclear Data File (ENSDF; Firestone 1996),
as well as the experimental nuclear masses (Audi and Wapstra 1995).
For theoretical masses,
currently the FRDM (M\"oller et al. 1995)
is favored. Microscopic information needed for the calculation of level
densities and $\alpha$+nucleus potentials are also taken from the FRDM, as
well as experimentally unknown ground state spins 
(M\"oller, Nix, \& Kratz 1997).

Finally, isobaric analog states $T^>=T^<+1=T^{\rm g.s.}+1$ are
explicitly considered in the new code (\cite{rauoak}). This replaces the
empirical constant suppression factor of the
$\gamma$-width used in
the WFHZ rates for self-conjugate nuclei.

For comparison, we give in Table \ref{vgl} a select set of {\it preliminary}
reaction rates obtained with the code NON-SMOKER (Rauscher and Thielemann 1998),
along with their counterparts in the TAT and WFHZ rate libraries.
These rates are identified (in the next section) as having influenced
the observed discrepancies in the stellar yields for the hybrid
models of $\S$ \ref{stel_ylds}. 

\section{Nuclear Origin of the Nucleosynthetic Differences}
\label{nucdif}

The nuclei which show greater than 20\% deviations in either the 15
M$_\odot$ or the 25 M$_\odot$ star are: $^{33}$S, $^{40}$Ar, $^{40}$K,
$^{44,46}$Ca, $^{45}$Sc, $^{50}$V, and $^{70}$Zn. Each
of these will now be discussed.  Quoted ratios
of stellar yields and reaction rates here will always be given as
those of TAT divided by WFHZ. The related reaction rates are given
in Tables \ref{vgl} and \ref{agoscn}.

\begin {figure}
\hskip 0.5 in \epsffile[0 0 612 612] {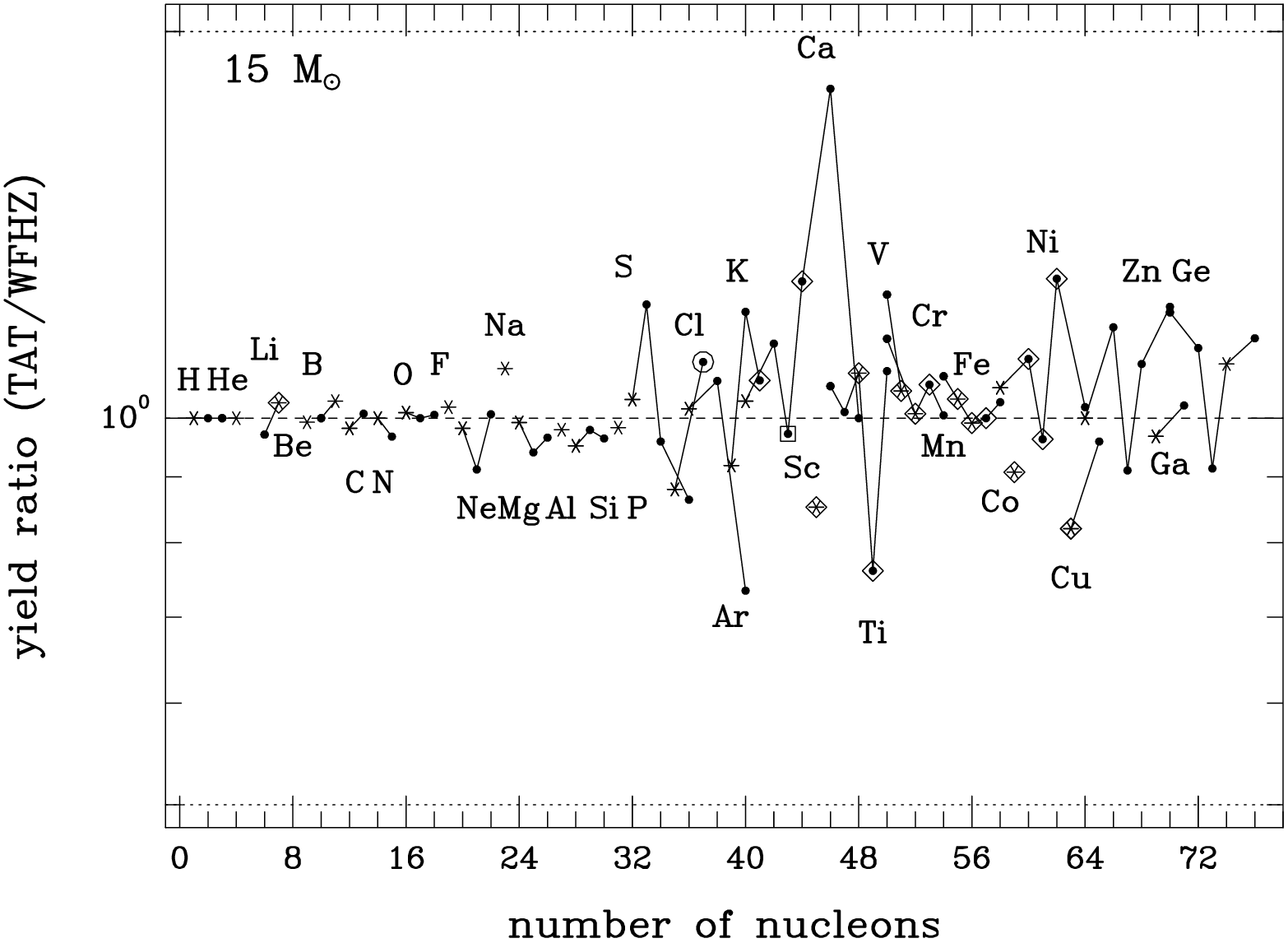}
\vskip -0.5 in
\caption {Comparison of nucleosynthesis using TAT rates 
and WFHZ rates (for A $\ge$ 28]. Plotted is the ratio of the final stellar yield (M\sun)
for a 15 M\sun \ supernova evolved from the main sequence and
calculated using TAT rates divided by the yield
calculated in an identically derived star using WFHZ rates. Below A =
28 the rate sets were identical. All models used a rate for 
$^{12}$C($\alpha,\gamma$)$^{16}$O from Caughlan \& Fowler (1988) multiplied 
by 1.7.}
\label {fig1}
\end{figure}

\begin {figure}
\hskip 0.5 in \epsffile[0 0 612 612] {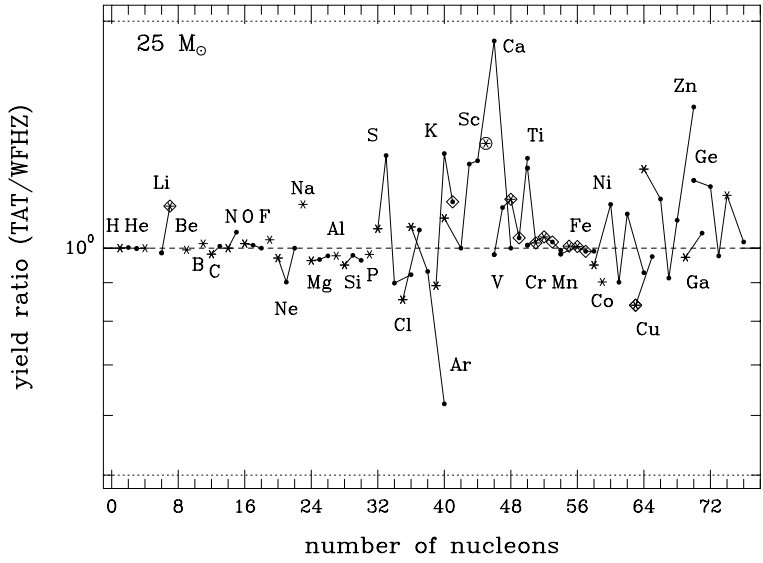}
\vskip -0.5 in
\caption {The same as Fig. \ref{fig1} but for a 25 M\sun \ supernova.}
\label {fig2}
\end{figure}

The nucleus $^{33}$S is predominantly made in explosive oxygen burning
(T$_9\sim 3.5$) 
and explosive neon burning (T$_9\sim 2.5$) (Thielemann \& Arnett 1985,
WW95). 
The main production and destruction reactions are
$^{32}$S(n,$\gamma$)$^{33}$S and
$^{33}$S(p,$\gamma$)$^{34}$Cl respectively. 
$^{30}$Si($\alpha$,n)$^{33}$S (and its inverse) can also compete,
depending on the neutron and $\alpha$-particle abundances and temperature.
It can even be the dominant channel for production or destruction. 
In the 15 M$_\odot$ and 25 M$_\odot$ stars, the $^{33}$S yield ratio 
(Y(TAT)/Y(WFHZ), Figures \ref{fig1} and \ref{fig2}) differed by
factors of 1.23 and 1.33, respectively. 
At a temperature of T$_9=3.5$ as shown in Table \ref{vgl}, 
the ratio of the production rates for  
$^{32}$S(n,$\gamma$)$^{33}$S is up by 
a factor of 1.3  and the destruction rate
$^{33}$S(p,$\gamma$)$^{34}$Cl is up by a factor of 2.0. However, 
$^{33}$S(n,$\alpha$)$^{30}$Si is down by 0.7.  
This combination, in addition to the particle fluxes at
these high temperatures, leads to the observed enhancement. It 
seems  the ($\alpha$,n) reaction works more in favor of destruction.

$^{40}$Ar and $^{40}$K are the result of neutron capture during 
hydrostatic helium, carbon, and to a lesser extent, neon burning. 
Since these two nuclei are linked by a common reaction flow, the
reasons for their observed discrepancies in the stellar yields are
likewise coupled. The $^{40}$Ar yield ratio is
down by factors of 0.73 and 0.62 while $^{40}$K
is up by a factor of 1.21 and 1.33, respectively, in the 15 and 25 M$_\odot$
stars.

In helium and carbon burning the neutron sources are
$^{22}$Ne($\alpha$,n)$^{25}$Mg and $^{13}$C($\alpha$,n)$^{16}$O.  Which
dominates depends on the stellar conditions and current state of
evolution.  At carbon burning temperatures ($\sim 1\times 10^9$K),
$^{40}$K is produced by $^{39}$K(n,$\gamma$)$^{40}$K and destroyed by
$^{40}$K(n,$\alpha$)$^{37}$Cl or $^{40}$K(n,p)$^{40}$Ar.  While the
ratio of the production reaction rates is slightly enhanced (by a
factor 1.2), the two destruction reaction ratios are suppressed by
factors of 0.77 and 0.4.  This leads to the enhancement of $^{40}$K by
factors of 1.21 and 1.33 seen in the 15 and 25 M$_\odot$ stars
respectively.  Similarly the ratio of the production rate,
$^{39}$Ar(n,$\gamma$)$^{40}$Ar, is near unity, but the production is
reduced by a factor 0.4 in the second channel $^{40}$K(n,p)$^{40}$Ar
mentioned above. This leads to the reduced production by factors of
0.73 and 0.62 seen in Figures \ref{fig1} and \ref{fig2} .

$^{46}$Ca is a product of neutron capture in hydrostatic helium, carbon and
neon burning. In the 15 and 25 M$_\odot$ models the yield ratios are
1.8 and 1.88 respectively.
In the chain $^{45}$Ca(n,$\gamma$)$^{46}$Ca(n,$\gamma$)$^{47}$Ca
the ratio of the producing (n,$\gamma$) reaction rates is higher by a
factor of 1.6, while the destruction reaction rate ratio is near unity.
This accounts for most of the alteration in the yield.

A special note should be made here in reference to the implementation
of experimental (n,$\gamma$) cross section data that lead to three of
the reaction rates given in Table \ref{vgl}. This accounts for some of
the differences seen in the neutron capture rates on $^{32}$S,
$^{39}$K, and $^{46}$Ca which affect the production of $^{33}$S,
$^{40}$K and $^{46}$Ca respectively.  In the WFHZ rate set, the
leading parameter in the fit to the theoretical (n,$\gamma$) rate
(``A'' in eq.\ 3, WFHZ, 1978) has been modified to give a reaction
rate at 30 keV that agrees with the suggested 30 keV cross section of
Bao \& K\"appeler (1987), while keeping the temperature dependence of
the original theoretical (n,$\gamma$) rate. The TAT set also made use
of these experimental rates, assuming however a dominant s-wave behavior which
leads to a constant rate as a function of temperature. Therefore both
implementations only agree at about 30 keV ($\equiv 3 \times 10^8$ K),
with the result that the TAT rate under(over)estimates the
(n,$\gamma$) rate at lower(higher) temperatures, consistent with the
$1/v$ behavior of neutron capture rates, although the deviation in
most cases is not large, often less than a factor of three, over a
wide range in temperature. This leads to the following different
neutron capture rates at the relevant temperatures in the WFHZ and TAT
rate sets, respectively: For $^{32}$S at $T_9$=3.5 the (n,$\gamma$)
rates are $5.4\times 10^5$ and $6.9\times 10^5$, for $^{39}$K at
$T_9$=1.0 they are $1.4\times 10^6$ and $1.7 \times 10^6$, and for
$^{46}$Ca at $T_9$=1.0 $7.8\times 10^5$ and $7.9\times 10^5$ (all
rates given in cm$^3$ s$^{-1}$ mole$^{-1}$).  Thus, the overproduction
of $^{33}$S and $^{40}$K, respectively, in the TAT models as compared
to the WFHZ models would be less pronounced with the same
implementation of the experimental rates.

$^{44}$Ca is dominantly produced as $^{44}$Ti in the $\alpha$-rich freeze-out
phase of explosive silicon burning. It is thus the product of the
$\alpha$-capture chain along  self-conjugate $\alpha$-nuclei
$^{28}$Si, $^{32}$S, $^{36}$Ar, $^{40}$Ca and $^{44}$Ti. 
In Table \ref{agoscn} 
we give the stellar ($\alpha,\gamma$) reaction rates on self-conjugate
nuclei from Mg to Ti (all evaluated at T$_9=2.5$)
as derived from the statistical model codes CRSEC, SMOKER, and NON-SMOKER,
respectively. 

For the ($\alpha,\gamma$) rates up to and including
$^{24}$Mg($\alpha,\gamma$)$^{28}$Si, the same experimental information
(CF88) has been used in both rate sets. The reason for the larger
SMOKER rates is the
neglect of isospin suppression effects for self-conjugate nuclei that 
were included in the CRSEC rates.
This effect is apparent in the new NON-SMOKER rates, which, at least
up to $^{36}$Ar, are in good agreement with those of CRSEC. The 
$^{24}$Mg($\alpha,\gamma$)$^{28}$Si 
reaction cross section has been measured (CF88) and the 
stellar rate (3.7 cm$^3$ s$^{-1}$ mole$^{-1}$ at $T_9$=2.5)
is in very good agreement with the new NON-SMOKER result 
(2.8 cm$^3$ s$^{-1}$ mole$^{-1}$). 
In the case of $^{40}$Ca($\alpha,\gamma$)$^{44}$Ti, experimental
data is also available (Cooperman, Shapiro \& Winkler 1977) but was not
used in the calculation of this cross section by either TAT or WFHZ. 
The experiment suggests a value 
in the temperature range of interest (T$_9\approx 2.5$) which
is about 4 times lower than the CRSEC rate, in excellent agreement with the
NON-SMOKER prediction, and in considerable disagreement with SMOKER
(Wiescher, private communication). 
All of this is encouraging, 
and indicates an area where modern laboratory measurements might lead to a
more reliable determination of some very important reaction rates.

It should be noted for the reactions that directly impact $^{44}$Ti synthesis
that their exact values 
will not affect abundances as long as temperatures are high enough 
to sustain an $(\alpha,\gamma)-(\gamma,\alpha)$-equilibrium, wherein
the abundances are only determined by Q-values and partition functions,
but the different rates will partially
enter during the charged particle freeze-out around 3$\times 10^9$K,
resulting in the enhancement of the $^{44}$Ti yield ratio by factors of
1.28 and 1.31, respectively, for the
15 and 25 M$_\odot$ models. This is particularly apparent on
consideration of the large disparity between the TAT and WFHZ
rates for the dominant production reaction 
($^{40}$Ca($\alpha,\gamma$)$^{44}$Ti, 
which differs by a factor of nearly 6), and the dominant destruction
reaction [$^{44}$Ti($\alpha$,p)$^{47}$V, which is identical between
the two sets and also shows no change in the preliminary NON-SMOKER calculation,
see Tables \ref{vgl} and \ref{agoscn}]. 
This is a nice example of where large
reaction rate differences cause much smaller abundance deviations
due to equilibria at high-temperatures in explosive burning and
only enter during the final freeze-out phase. 

Several other reaction rates of importance to $^{44}$Ti synthesis have been 
identified by The et al. (1998), who used the TAT rate set in their
studies. One such reaction not previously mentioned is 
$^{45}$V(p,$\gamma$)$^{46}$Cr which is on the border of a QSE cluster that
affects $^{44}$Ti. At 
temperatures where the material is falling out of equilibrium (T$_9\approx 2.5$)
this rate becomes  important, and the TAT rate is in severe
disagreement with the WFHZ rate, with a reaction ratio of TAT/WFHZ=0.016.
This had no impact on the present calculations, because
$^{46}$Cr was not included in our networks. The et al. also point
out that for a composition with a non-zero neutron excess ($\eta \ge 0.0004$),
this reaction rate does not affect $^{44}$Ti synthesis.

$^{45}$Sc is made in two complementary ways via the radioactive progenitor
$^{45}$Ti in explosive oxygen and silicon burning, and directly during neon
and carbon burning. In the 25M$_\odot$ star, and using the TAT rate set, $^{45}$Sc
was made in equal proportions as itself and radioactive $^{45}$Ca. 
When using the
WFHZ rare set, it was made (predominantly) as itself. 
In the 15 M$_\odot$ star the production via $^{45}$Ti in explosive oxygen and 
silicon burning dominates. The production is reduced by a factor of
0.85 in the yield comparison TAT/WFHZ. This occurs via the reaction sequence  
$^{42}$Ca($\alpha$,n)$^{45}$Ti(n,p)$^{45}$Sc 
and $^{43}$Ca(p,$\gamma$)$^{44}$Sc(p,$\gamma$)$^{45}$Ti.
In the first sequence, the (n,p)-reaction is enhanced by a factor
1.4 (i.e. enhanced $^{45}$Ti destruction).  The first rate in the proton
capture sequence, which controls the flow, is reduced by a factor 0.9.
In combination this lead to the reduction of a factor of 0.85. 

$^{50}$V is made in explosive and hydrostatic neon burning, coupled with
some neutron processing. The destruction rate $^{50}$V(n,$\gamma$)$^{51}$V
is reduced by a factor of 0.75, thus leading to an enhancement of factors
1.25 and 1.28 respectively for the 15 and 25 M$_\odot$ stars.

The heavy nucleus $^{70}$Zn is made by neutron
capture in He, C, and Ne-burning, i.e., the $s$-process. It's
abundance variation can be explained by changes in neutron capture
cross sections. The reaction chain $^{69}$Zn(n,$\gamma$)$^{70}$Zn(n,$\gamma$)$^{71}$Zn
has the first rate increased by a factor of 1.97, the second by
1.7. This lead to a net gain of a factor 1.22.  

\section{Summary and Conclusions}

We have reviewed the theoretical nuclear reaction rates currently used
for modern calculations of stellar nucleosynthesis in the intermediate
mass range (magnesium through krypton) and have explored the
sensitivity of standard calculations to those rates. Both sets, the
one of Thielemann et al. (TAT) and the one of Woosley et al. (WFHZ),
are derived from Hauser-Feshbach theory. We discussed the underlying
assumptions in each rate set, providing much previously unpublished
information. We compared them to one another and to experiment,
elucidating the differences, and explored the nucleosynthesis obtained
using each rate set in otherwise identical models for the evolution
and explosion of 15 and 25 M\sun \ supernovae.

\begin{figure}
\epsfig{file=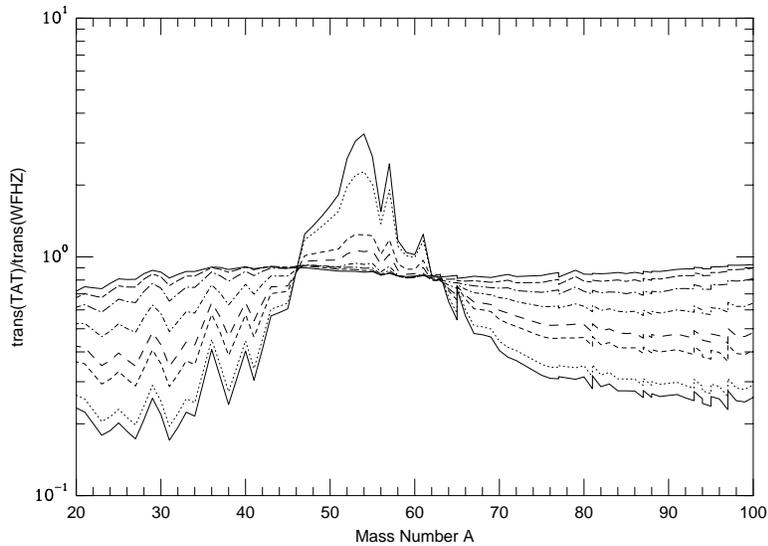,width=9cm,angle=0.}
\caption{\label{s_trans_n} Ratios of neutron transmission functions 
obtained with the TAT
potentials and the equivalent square well description of WFHZ
for a set of stable nuclei with varying mass number A. Shown are
the ratios for s-wave neutrons. Different line styles denote different
bombarding energies of the target-projectile system in the
center of mass frame: 0.01 MeV (solid), 0.1 MeV (dotted),
1.0 MeV (short dashes), 2.0 MeV (long dashes), 5.0 MeV (dot --
short dash), 10.0 MeV (dot -- long dash), 15.0 MeV (short dash -- long dash),
20.0 MeV (solid).}
\end{figure}

\begin{figure}
\epsfig{file=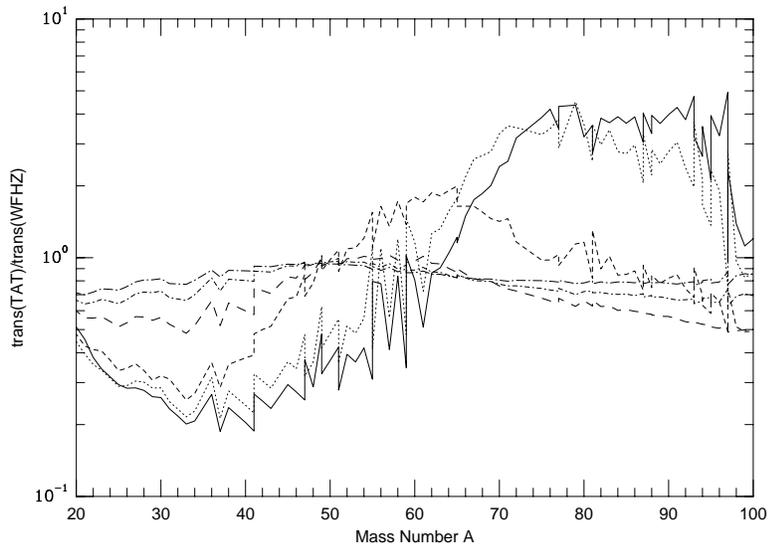,width=9cm,angle=0.}
\caption{\label{s_trans_p} Same as Figure \ref{s_trans_n} but for
ratios of s-wave proton transmission functions. 
The line styles denote different energies:
1.0 MeV (solid), 2.0 MeV
(dotted), 5.0 MeV (short dashes), 10.0 MeV (long dashes), 15.0 MeV (dot --
short dash), 20.0 MeV (dot -- long dash).}
\end{figure}

\begin{figure}
\epsfig{file=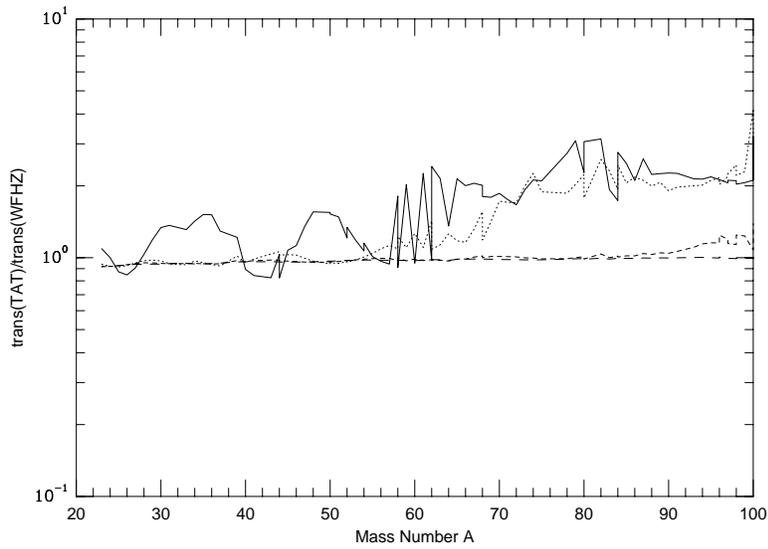,width=9cm,angle=0.}
\caption{\label{s_trans_a} Same as Figure \ref{s_trans_n} but for
ratios of s-wave alpha-particle transmission functions.
Line styles denote different energies:
5.0 MeV (solid), 10.0 MeV
(dotted), 15.0 MeV (short dashes), 20.0 MeV (long dashes).}
\end{figure}

\begin{figure}
\epsfig{file=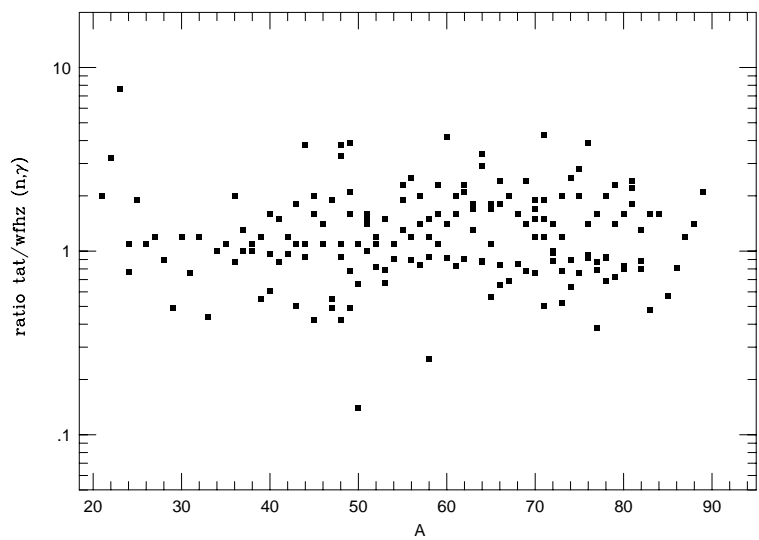,width=9cm,angle=90.}
\caption{\label{ng} Ratios of TAT over WFHZ (n,$\gamma$) rates
evaluated at T$_9=2.5$.}
\end{figure}

\begin{figure}
\epsfig{file=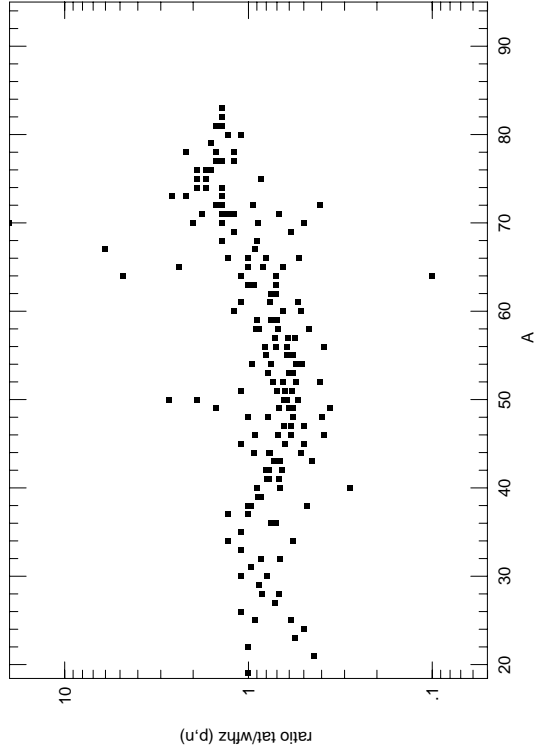,width=9cm,angle=90.}
\caption{\label{pn} Ratios of TAT over WFHZ (p,n) rates evaluated
at T$_9=2.5$.The extreme values at A=64 and 70 (0.1 and 20 respectively),
give an indication of the differences that arise when calculating
theoretical reaction rates using nuclear data far from stability.
The two cases are for (p,n) rate ratios on targets $^{64}$Ga and $^{70}$Se 
(each 5 and 4 units from stability
towards the proton drip line) which have (p,n) Q-values that differ by 
97 and 540 keV (respectively)
in the two rate sets. Also, in the first case the spin of the compound nucleus
differs by 1, in the second case it differs by 1 for the compound nucleus
and 2 for the final nucleus. This leads to different channel openings
due to Q-values and angular momentum barriers and shows how crucial
mass and ground state properties are far from stability.}
\end{figure}

\begin{figure}
\epsfig{file=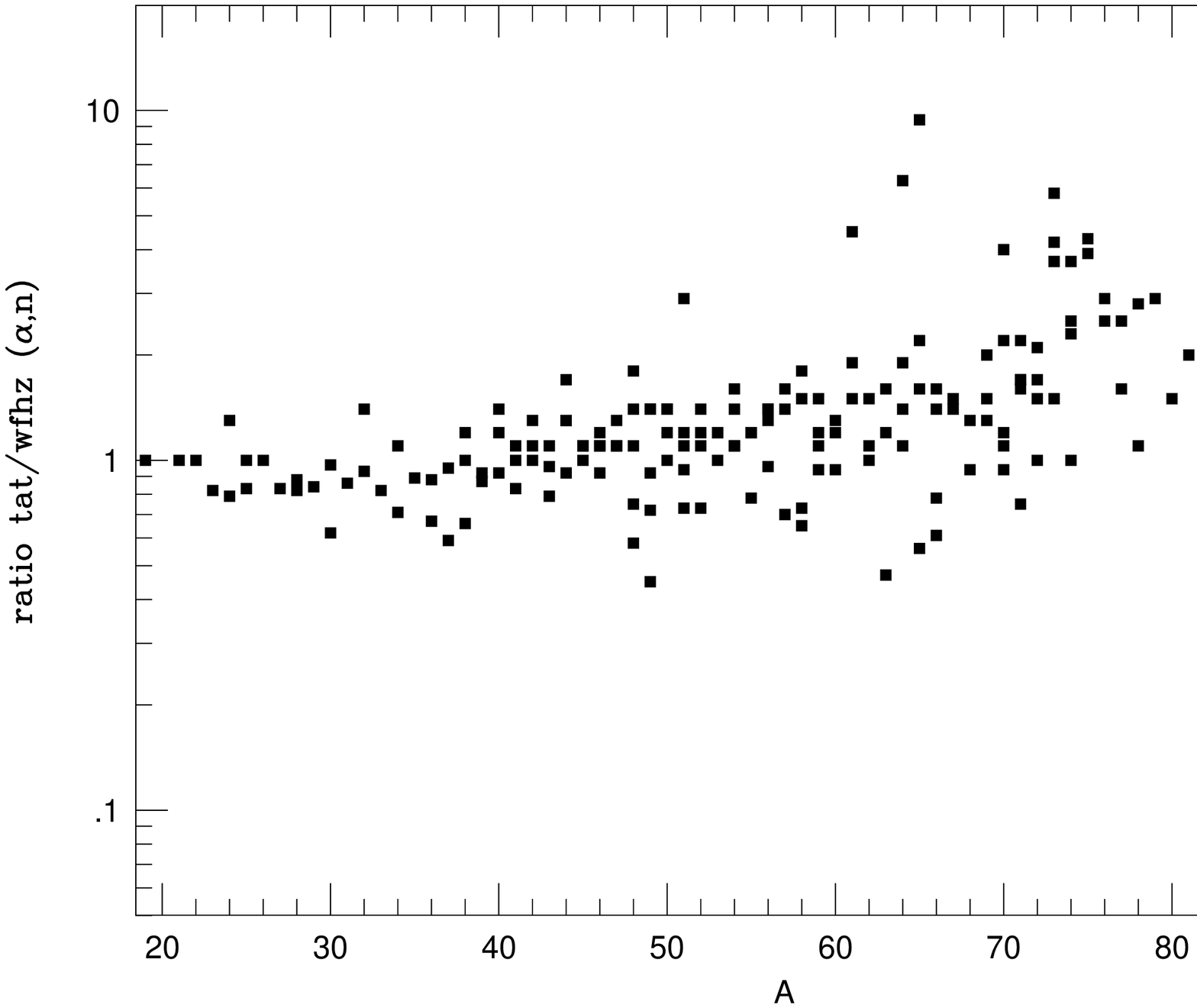,width=9cm,angle=90.}
\caption{\label{an}Ratios of TAT over WFHZ ($\alpha$,n) rates
evaluated at T$_9=2.5$.}
\end{figure}

\begin{figure}
\epsfig{file=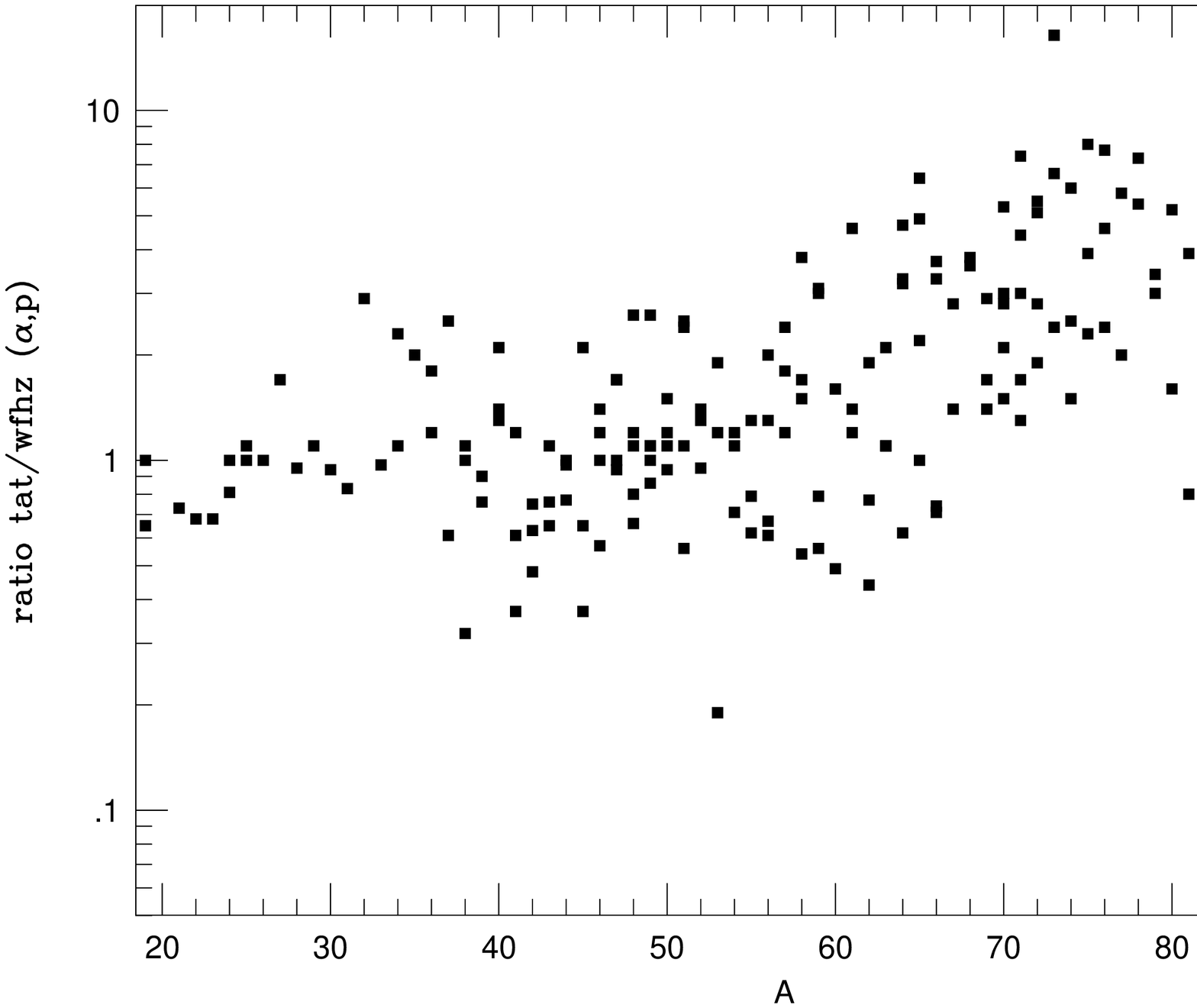,width=9cm,angle=90.}
\caption{\label{ap}Ratios of TAT over WFHZ ($\alpha$,p) rates
evaluated at T$_9=2.5$.}
\end{figure}

Overall, the two 
rate sets are similar. Typical differences at
astrophysically interesting temperatures are less than a factor of two
(Figures \ref{ng} - \ref{ap}). There are individual cases, however, where the
difference exceeds a factor of 10. Some of the larger differences
occur for reactions where scarce experimental information is available
and different assumptions were made regarding the photon transmission
function, for example, ($\alpha,\gamma$) reactions on Z = N 
nuclei. Different assumptions were also made about the particle
transmission functions, nuclear partition functions, and level
densities. More modern and complete data used in the TAT rates makes
them superior in cases where the partition function is
important. Assumptions regarding the particle transmission functions
seem less important (see, however, Rauscher 1998 and Somorjai et al. 1998
with respect to alpha transmission functions for heavy nuclei).  
WFHZ used an equivalent square well with
empirical reflection factors; TAT used a more detailed optical
model. Given the quite different values for, e.g., the neutron and
proton transmission function (Figures \ref{s_trans_n} and \ref{s_trans_p}), 
it is perhaps surprising
that the rates (e.g., for (p,n) reactions, Fig. \ref{pn}) differ so little.
This is because the relevant temperatures for explosive burning are
high. For incident particles in the Gamow window, the deviations in
the particle transmission functions are typically smaller than a
factor of two. In addition, higher partial waves (not shown in 
Fig. \ref{s_trans_p})
contribute.  A comparison of rates at a lower temperature would have
revealed larger discrepancies.

\begin{figure}
\epsfig{file=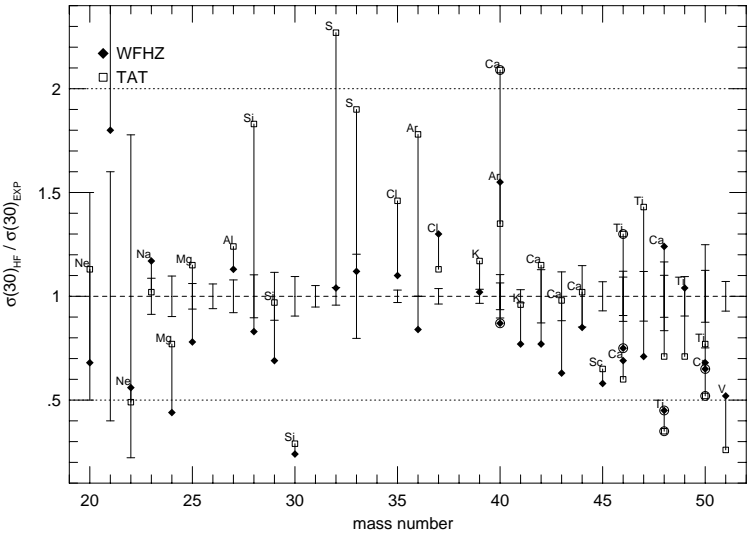,width=9cm,angle=0.}
\caption{\label{ngcxl}30 keV neutron-capture cross section
ratios vs. mass number (20$\le$ A $\le$ 50);
Theory (WFHZ: filled diamonds, TAT: open squares) divided
by experiment. Error bar values (error/cross section) are shown for each
comparison ratio. Ratios for (n,$\gamma$) cross sections on common
targets are connected by a solid vertical line.
For reactions proceeding on targets with a common mass number,
the second pair is circled (eg. $^{40}$Ca).}
\end{figure}

\begin{figure}
\epsfig{file=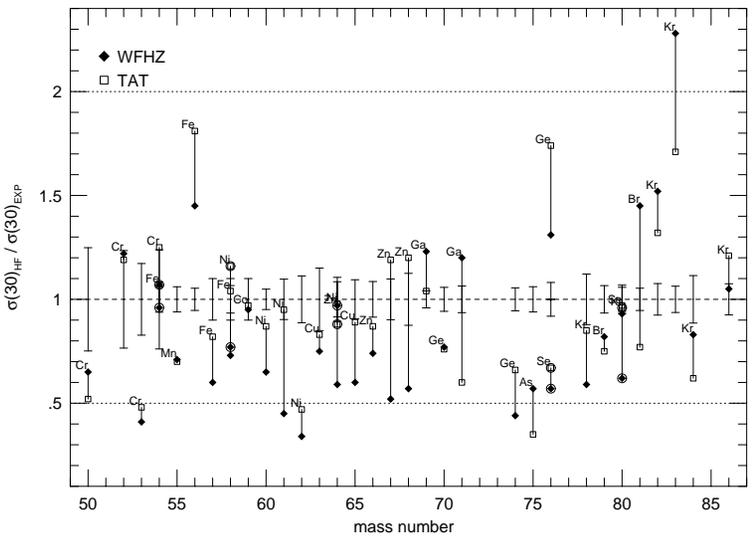,width=9cm,angle=0.}
\caption{\label{ngcxh}
Same as Figure \ref{ngcxl}, but for 50 $\le$ A $\le$ 86.}
\end{figure}

\begin{figure}
\epsfig{file=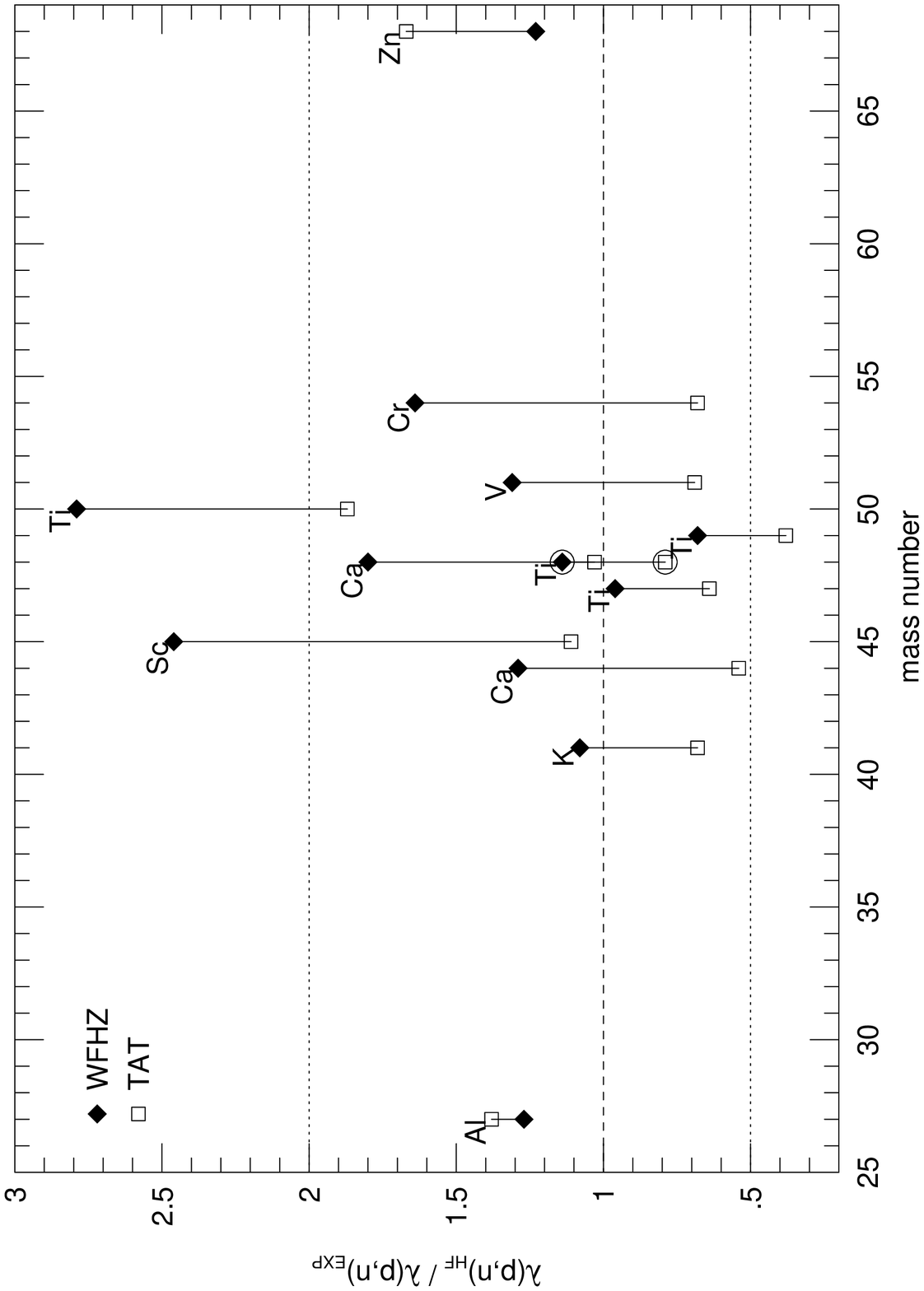,width=9cm,angle=0.}
\caption{\label{pncx}Proton-induced (p,n) reaction rate
ratios vs. mass number A;
Theory (WFHZ: filled diamonds,
TAT: open squares) divided by experiment. Ratios for (p,n)-reactions on
common targets are connected by a solid vertical line.
For reactions proceeding on
targets with a common mass number, the second pair is circled (eg. $^{48}$Ti).}
\end{figure}

\begin{figure}
\epsfig{file=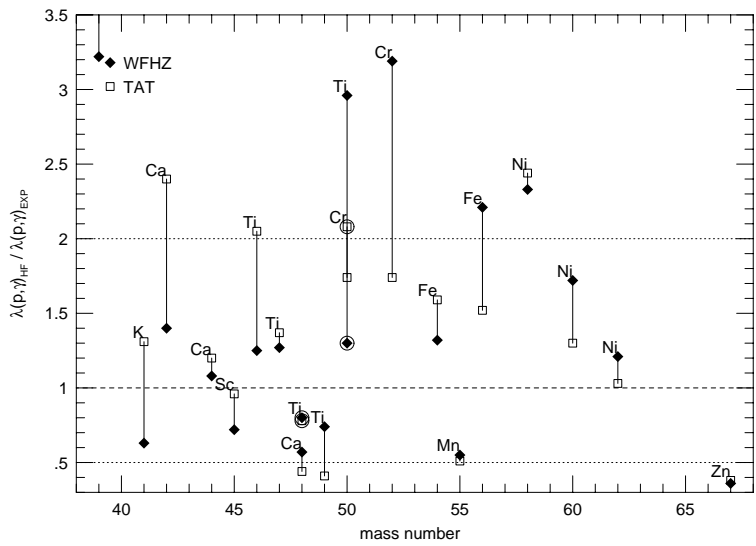,width=9cm,angle=0.}
\caption{\label{pgcx}Same as Figure \ref{pncx} but for 
(p,$\gamma$) reaction ratios.}
\end{figure}

\begin{figure}
\epsfig{file=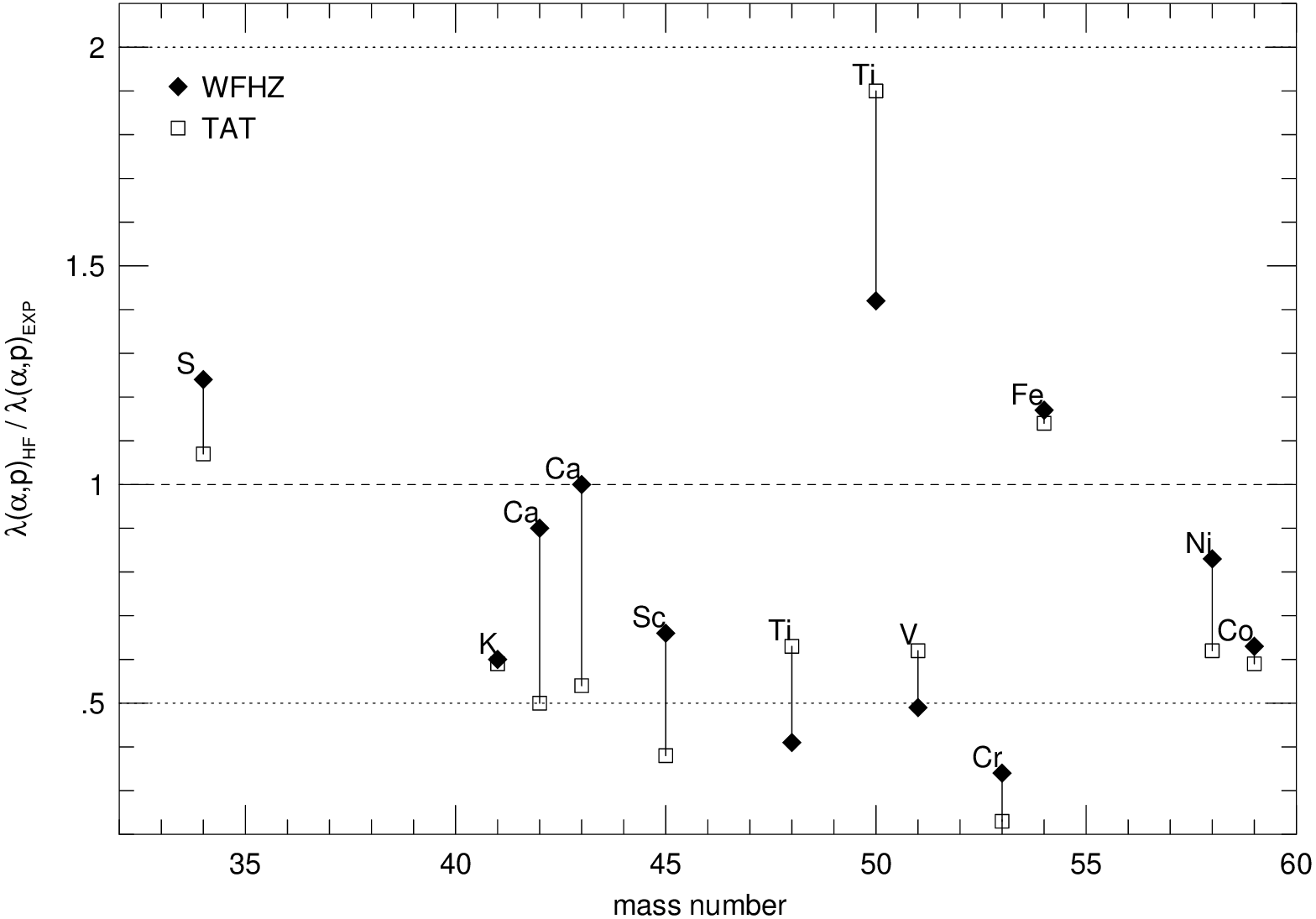,width=9cm,angle=0.}
\caption{\label{apcx}Same as Figure \ref{pncx} but for 
($\alpha$,p) reaction ratios.}
\end{figure}

\begin{figure}
\epsfig{file=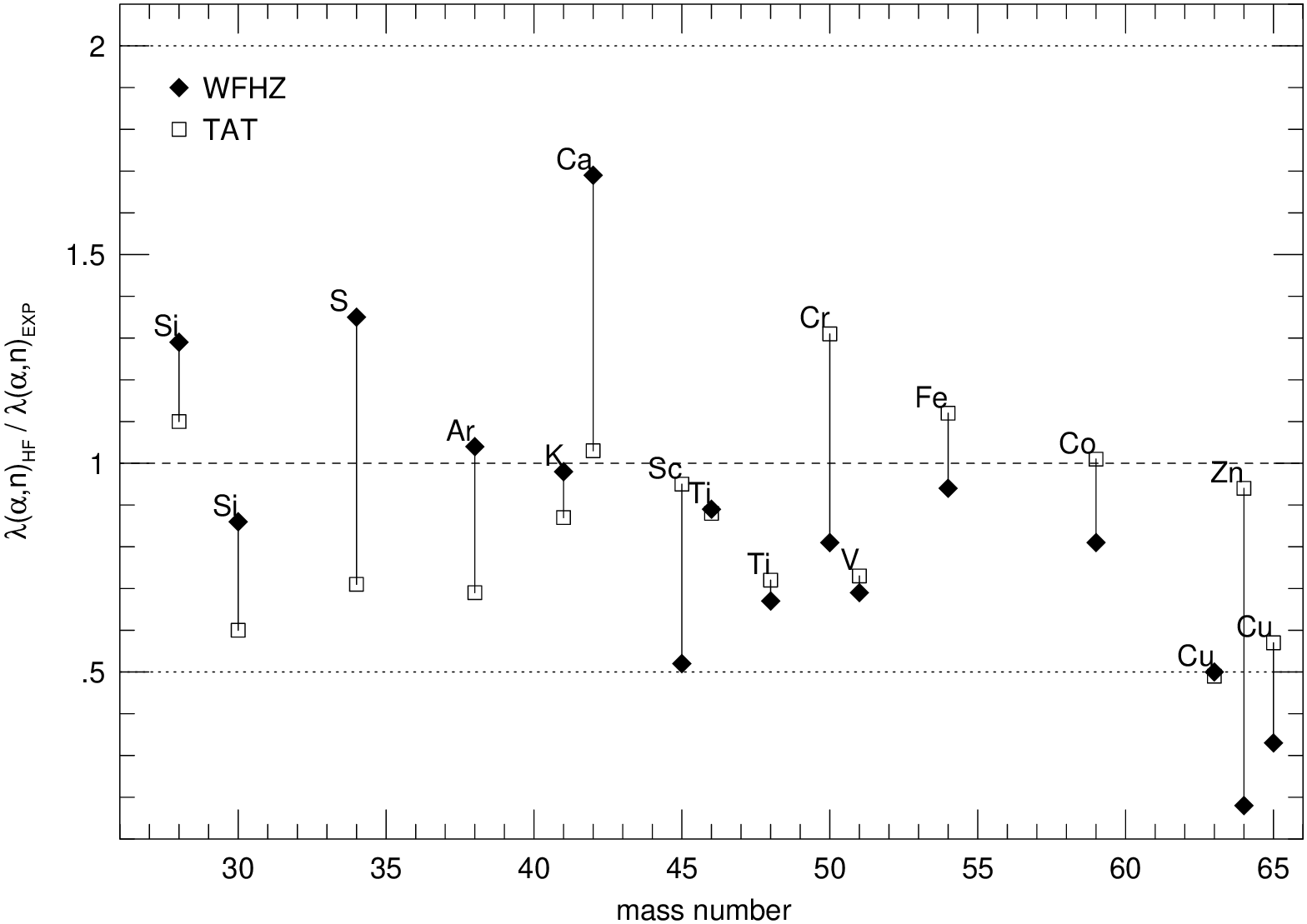,width=9cm,angle=0.}
\caption{\label{ancx}Same as Figure \ref{pncx} but for 
($\alpha$,n) reaction ratios.}
\end{figure}

Compared to experiment, both sets of theoretical rates give similar
agreement, typically to a factor of two (Figures \ref{ngcxl} - \ref{ancx}).
The standard deviations
between the two theoretical sets and data for (n,$\gamma$) cross
sections - which measure how well the photon transmission function is
calculated - are virtually identical, 0.58 for WFHZ/EXP and 0.64 for TAT/EXP
($\S$4.2.2) and both sets predict a mean that is equally close to
experiment, TAT/EXP is higher (1.08), WFHZ/EXP is lower (0.91). For
charged particles - which are more sensitive to particle transmission
functions - the TAT calculations are slightly superior, with standard
deviations over the entire range of experimentally know reaction rate
data being 0.81 and 0.65 [for (p,$\gamma$) rates] and
0.62 and 0.47 [(p,n) rates] for WFHZ/EXP and TAT/EXP respectively.
Other comparisons are given in Table \ref{thvexp} and $\S$4.2.2.

In summary, the two rate sets have comparable merit when compared to
experiment. All the authors of this paper agree that the new rate set
- the ``NON-SMOKER" set, will be preferable to both TAT and WFHZ and,
when they become available, will be adopted by both groups for
future work.

When the two current rate sets are included in otherwise identical
stellar models we find that the nucleosynthesis, again with rare and
occasionally interesting exceptions, is not greatly
changed. Individual exceptions are discussed in $\S$5.1. For example, 
only about a
dozen (out of 70) stable isotopes in the mass range 12 to 70 have
nucleosynthesis that differ by over 20\% in two supernovae of 
15 M\sun \ that use the same rate for
$^{12}$C($\alpha,\gamma$)$^{16}$O (Table \ref{new15vstnh96}).
It can, however, be noticed that most of these isotopes - with one exception
$^{44}$Ti - are products of hydrostatic burning where individual reaction
flows are governed by the cross sections involved.
Nevertheless, none differ by more than a factor of 1.7. Given the significantly
larger differences that exist in individual reaction rates, one may
wonder at the robust nature of the final nucleosynthesis. We see three
major causes.

First, as the star burns and becomes hotter, the nuclear flow follows
the valley of beta-stability making heavier nuclei as it goes. In
doing so, it follows the path of least resistance - those reactions
having the largest cross section for a nucleon or $\alpha$-particle
reacting with a given nucleus. These large cross sections are
reasonably well replicated by any calculation, normalized to
experiment, that treats the Coulomb barrier and photon transmission
function approximately correctly. Large differences may exist in rate
factors for reactions that are in competition, especially a small
channel in the presence of a large one, but these small channels are
frequently negligible, at least for the major abundances while they can cause
larger differences when one is interested in the abundances of trace isotopes.

If one is however interested in accurate abundance predictions resulting from 
these smaller flows in hydrostatic burning stages, this can in most cases
only be obtained by improving the reliability of the 
cross sections (and reaction rates) that determine these weak flows
on light and intermediate mass nuclei. As new experimental information
becomes available, a continuous improvement is therefore highly
warranted (see e.g. the most recent compilation of experimental rates, 
NACRE, Angulo et al. 1999).

Second, beyond oxygen burning, which is to say for nuclei heavier than
calcium, nucleosynthesis increasingly occurs in a state of full or
partial nuclear statistical equilibrium. There the abundances are
given by binding energies and partition functions. So long as the
``freeze-out" is sufficiently rapid, individual rates are not so
important.

Third, the reaction rates varied here were only those theoretical
values from Hauser-Feshbach calculations for intermediate mass nuclei,
i.e., nuclei heavier than magnesium. The really critical reaction
rates are, for the most part, those below magnesium. These reactions,
like e.g., $^{12}$C($\alpha,\gamma$)$^{16}$O, govern the energy
generation, major nucleosynthesis, and neutron exposure in the
star. The rest are perturbations on these dominant flows.

This is not to say, however, that the nuclear and stellar details of
heavy element synthesis are now well understood. Differences in the
{\sl stellar model} may account, not just for 20\% variation, but {\sl
orders of magnitude} (Table \ref{new15vstnh96}; column TNH). That is, uncertainty in
stellar physics - especially the treatment of convection and how it
is coupled (or not coupled) to the nuclear network - accounts for most
of the differences in current nucleosynthesis calculations - provided
such calculations use the same nuclear reaction rates below magnesium.

Even in a perfect stellar model though, there will still be
interesting nuclear physics issues. Stellar nucleosynthesis is
becoming a mature field rich with diverse and highly detailed
observational data. The ``factor of two" accuracy that was adequate in
the past may not do justice to the observations of the future. There
are many individual cases where the nuclear physics uncertainty is
still unacceptably large. We would close by just pointing out two of
them. 

The suppression of radiative capture reactions into self-conjugate
(isospin zero) nuclei is very uncertain. Past Hauser-Feshbach calculations
have adopted empirical factors for this suppression
($\S$4.2.1). Alpha-capture reactions, like $^{24}$Mg($\alpha,\gamma$)$^{28}$Si,
$^{28}$Si($\alpha,\gamma$)$^{32}$S, ...,
$^{44}$Ti($\alpha,\gamma$)$^{48}$Cr, are very important to
nucleosynthesis in oxygen and silicon burning. The reaction
$^{40}$Ca($\alpha,\gamma$)$^{44}$Ti also directly affects the synthesis
of $^{44}$Ti. Modern accurate determinations of most of the reaction
rates are missing (as well as (p,$\gamma$) reactions into the same
nuclei). Measurements here would be most welcome.

The Hauser-Feshbach rates are also only as good as the local
experimental rates to which the necessary parameters of the
calculation are calibrated. In that regard we would point out the near
absence of charged particle reaction rate data for A $\gtaprx$ 70. Charged
particle reactions are important, especially on unstable nuclei, at
significantly higher atomic weights in the $r$-process (Hoffman,
Woosley \& Qian 1997) and in the $p$-process (Rayet et al. 1995).

{\bf Acknowledgments.}

The authors appreciate many valuable conversations with Ken Nomoto
regarding stellar evolution and explosive nucleosynthesis.
This work has been supported by the National Science Foundation through grants
AST-96-17494 (Steward Observatory, UA), 
AST-96-17161 and AST-97-31569 (Lick Observatory, UCSC), and
PHY74-07194 (Institute for Theoretical Physics, UCSB). 
S. E. Woosley was supported at the 
Max Plank Institute for Astrophysics (Garching) under a grant from 
the Alexander von Humboldt Foundation.
F.-K. Thielemann and T. Rauscher were supported in Basel by
Swiss Nationalfonds Grants (20-47252.96 \& 2000-53798.98), 
and T. Rauscher gratefully
acknowledges support through an APART fellowship from the 
Austrian Academy of Sciences.

\clearpage

\epsfxsize = 10. true cm

\clearpage

\epsfxsize = 10. true cm

\clearpage

\clearpage

\clearpage

\clearpage

\clearpage

%

\clearpage

\clearpage

\clearpage

\clearpage

%

\clearpage

\clearpage

\clearpage

\clearpage

\clearpage

\clearpage

\clearpage

%
%
%
%

\clearpage

\begin{deluxetable}{l c c c c c c }
\tablewidth{0pt}
\tablecaption{\label{ntwks}Nuclear Reaction Networks Employed}
\tablehead{
\colhead{Z}& \colhead{A$_{min}^a$}& \colhead{A$_{max}$}&
\colhead{A$_{min}^b$}& \colhead{A$_{max}$}&
\colhead{A$_{min}^c$}& \colhead{A$_{max}$}}
\startdata
 H & 1  & 3  & 1  & 3  & 1  & 3  \nl
He & 3  & 4  & 3  & 4  & 3  & 4  \nl
Li & 6  & 8  & 6  & 8  & 6  & 8  \nl
Be & 7  & 9  & 7  & 9  & 7  & 10 \nl
B  & 8  & 11 & 8  & 11 & 10 & 12 \nl
C  & 11 & 14 & 11 & 14 & 11 & 14 \nl
N  & 13 & 15 & 13 & 15 & 13 & 17 \nl
O  & 15 & 18 & 15 & 18 & 14 & 20 \nl
F  & 17 & 20 & 17 & 19 & 17 & 21 \nl
Ne & 19 & 23 & 19 & 23 & 18 & 25 \nl
Na & 21 & 25 & 22 & 24 & 21 & 26 \nl
Mg & 23 & 28 & 23 & 27 & 22 & 28 \nl
Al & 25 & 30 & 25 & 28 & 24 & 30 \nl
Si & 27 & 32 & 27 & 32 & 26 & 33 \nl
P  & 29 & 34 & 29 & 34 & 27 & 35 \nl
S  & 31 & 38 & 31 & 37 & 29 & 38 \nl
Cl & 33 & 41 & 33 & 38 & 31 & 39 \nl
Ar & 36 & 44 & 36 & 41 & 33 & 44 \nl
K  & 37 & 46 & 37 & 42 & 35 & 46 \nl
Ca & 40 & 50 & 40 & 49 & 37 & 49 \nl
Sc & 41 & 51 & 41 & 50 & 40 & 50 \nl
Ti & 43 & 52 & 44 & 51 & 42 & 52 \nl
 V & 45 & 54 & 45 & 52 & 44 & 54 \nl
Cr & 47 & 56 & 48 & 55 & 46 & 56 \nl
Mn & 49 & 59 & 51 & 57 & 48 & 58 \nl
Fe & 51 & 64 & 52 & 61 & 50 & 62 \nl
Co & 54 & 65 & 55 & 62 & 52 & 63 \nl
Ni & 55 & 68 & 56 & 65 & 54 & 67 \nl
Cu & 57 & 71 & 57 & 66 & 57 & 69 \nl
Zn & 59 & 74 & 60 & 69 & 59 & 72 \nl
Ga & 61 & 77 & 61 & 70 & 61 & 74 \nl
Ge & 64 & 80 & 64 & 71 & 63 & 78 \nl
As & 66 & 83 &  - &  - &  - &  - \nl
Se & 68 & 86 &  - &  - &  - &  - \nl
Br & 70 & 89 &  - &  - &  - &  - \nl
Kr & 72 & 90 &  - &  - &  - &  - \nl
\tablenotetext{a}{This study.}
\tablenotetext{b}{Woosley \& Weaver 1995.}
\tablenotetext{c}{Thielemann, Nomoto, \& Hashimoto, 1996.}
\enddata
\end{deluxetable}

\clearpage

\begin{deluxetable}{l c c c c c l}
\tablewidth{0pt}
\tablecaption{\label{new15vstnh96}15 M\sun \ Stellar Yields (in M\sun)
\ \ $^{12}$C($\alpha,\gamma$)=2.35*CF88}
\tablehead{
\colhead{$^{\rm A}$Z}&
\colhead{WFHZ}& 
\colhead{TAT}& 
\colhead{WFHZ/TAT}&
\colhead{TNH}&  
\colhead{TNH/TAT}&
\colhead{process(es)$^a$}
}
\startdata
 $^{1}$H   & 6.96     & 6.96     &   1.00 & ---      &  ---  & BB\nl
 $^{4}$He  & 5.21     & 5.21     &   1.00 & 1.83     &  0.351& BB,L*,H\nl
 $^{12}$C  & 1.38(-1) & 1.36(-1) &   1.01 & 8.33(-2) &  0.611& L*,He\nl
 $^{14}$N  & 5.27(-2) & 5.27(-2) &   1.00 & 5.37(-3) &  0.102& L*,H\nl
 $^{16}$O  & 6.31(-1) & 6.35(-1) &   0.99 & 4.23(-1) &  0.666& He\nl
 $^{18}$O  & 6.06(-3) & 6.06(-3) &   1.00 & 1.35(-2) &  2.228& He\nl
 $^{19}$F  & 6.38(-6) & 6.39(-6) &   1.00 & 2.67(-11)&  ---  & $\nu$,He\nl
 $^{20}$Ne & 2.60(-2) & 2.57(-2) &   1.01 & 2.83(-2) &  1.102& C\nl
 $^{21}$Ne & 1.11(-4) & 1.09(-4) &   1.02 & 4.53(-5) &  0.416& C\nl
 $^{22}$Ne & 7.48(-3) & 7.52(-3) &   1.00 & 1.26(-2) &  1.676& He\nl
 $^{23}$Na & 1.02(-3) & 1.03(-3) &   0.99 & 2.09(-4) &  0.202& C,Ne,H\nl
 $^{24}$Mg & 1.03(-2) & 1.21(-2) &   0.85 & 4.20(-2) &  3.483& C,Ne\nl
 $^{25}$Mg & 1.91(-3) & 1.92(-3) &   1.00 & 3.46(-3) &  1.803& C,Ne\nl
 $^{26}$Mg & 2.68(-3) & 2.67(-3) &   1.00 & 2.52(-3) &  0.945& C,Ne\nl
 $^{27}$Al & 1.33(-3) & 1.39(-3) &   0.95 & 5.55(-3) &  3.999& C,Ne\nl
 $^{28}$Si & 1.93(-1) & 1.70(-1) &   1.14 & 6.52(-2) &  0.384& xO,O\nl
 $^{29}$Si & 1.16(-3) & 1.22(-3) &   0.95 & 4.40(-3) &  3.615& C,Ne\nl
 $^{30}$Si & 6.05(-4) & 5.92(-4) &   1.02 & 4.91(-3) &  8.326& C,Ne\nl
 $^{31}$P  & 1.06(-3) & 9.79(-4) &   1.08 & 8.67(-4) &  0.886& C,Ne\nl
 $^{32}$S  & 1.65(-1) & 1.84(-1) &   0.90 & 2.16(-2) &  0.117& xO,O\nl
 $^{33}$S  & 1.18(-3) & 1.60(-3) &   0.74 & 9.31(-5) &  0.058& xO,xNe\nl
 $^{34}$S  & 7.70(-3) & 8.65(-3) &   0.89 & 1.09(-3) &  0.126& xO,O\nl
 $^{36}$S  & 3.22(-6) & 3.14(-6) &   1.03 & 5.27(-7) &  0.168& He(s),C,Ne\nl
 $^{35}$Cl & 1.15(-3) & 8.83(-4) &   1.30 & 5.46(-5) &  0.062& He(s),xO,xNe,$\nu$\nl
 $^{37}$Cl & 3.38(-4) & 4.17(-4) &   0.81 & 5.62(-6) &  0.013& xO,xNe\nl
 $^{36}$Ar & 4.71(-2) & 4.73(-2) &   1.00 & 3.49(-3) &  0.074& xO,O\nl
 $^{38}$Ar & 1.39(-2) & 1.54(-2) &   0.90 & 3.26(-4) &  0.021& xO,O\nl
 $^{40}$Ar & 1.11(-6) & 9.58(-7) &   1.16 & 4.12(-9) &  0.004& He(s),C,Ne\nl
 $^{39}$K  & 1.68(-3) & 1.18(-3) &   1.41 & 1.62(-5) &  0.014& xO,O,$\nu$\nl
 $^{40}$K  & 2.24(-6) & 1.73(-6) &   1.30 & 7.04(-9) &  0.004& He(s),C,Ne\nl
 $^{41}$K  & 2.45(-4) & 2.49(-4) &   0.98 & 1.25(-6) &  0.005& xO\nl
 $^{40}$Ca & 4.00(-2) & 3.84(-2) &   1.04 & 3.03(-3) &  0.079& xO,O\nl
 $^{42}$Ca & 7.63(-4) & 6.93(-4) &   1.10 & 6.98(-6) &  0.010& xO\nl
 $^{43}$Ca & 9.37(-6) & 8.47(-6) &   1.11 & 1.21(-6) &  0.143& C,Ne\nl
 $^{44}$Ca & 8.00(-5) & 9.36(-5) &   0.86 & 7.19(-5) &  0.768& $\alpha$,Ia-det\nl
 $^{46}$Ca & 7.38(-8) & 1.21(-7) &   0.61 & 8.50(-12)&  ---  & He(s),C,Ne\nl
 $^{48}$Ca & 1.75(-6) & 1.75(-6) &   1.00 & 6.71(-16)&  ---  & nse-Ia-MCh\nl
 $^{45}$Sc & 3.55(-6) & 2.91(-6) &   1.22 & 4.66(-8) &  0.016& $\alpha$,xO,xSi,C,Ne,$\nu$\nl
 $^{46}$Ti & 1.81(-4) & 1.90(-4) &   0.95 & 2.72(-6) &  0.014& xO,Ia-det\nl
 $^{47}$Ti & 1.02(-5) & 1.08(-5) &   0.94 & 4.67(-6) &  0.432& xO,xSi,Ia-det\nl
 $^{48}$Ti & 1.56(-4) & 1.72(-4) &   0.91 & 1.27(-4) &  0.738& xSi,Ia-det\nl
 $^{49}$Ti & 1.49(-5) & 1.18(-5) &   1.26 & 4.15(-6) &  0.352& xSi\nl
 $^{50}$Ti & 2.77(-6) & 2.83(-6) &   0.98 & 4.59(-10)&  ---  & nse-Ia-MCh\nl
 $^{50}$V  & 1.06(-7) & 1.34(-7) &   0.79 & 3.51(-10)&  0.003& C,Ne,xNe,xO\nl
 $^{51}$V  & 4.36(-5) & 5.90(-5) &   0.74 & 1.01(-5) &  0.171& $\alpha$,Ia-det,xSi,xO,$\nu$\nl
 $^{50}$Cr & 2.31(-4) & 2.83(-4) &   0.81 & 3.85(-5) &  0.136& xSi,xO,$\alpha$,Ia-det\nl
 $^{52}$Cr & 1.37(-3) & 1.43(-3) &   0.96 & 8.24(-4) &  0.576& xSi,$\alpha$,Ia-det\nl
 $^{53}$Cr & 1.70(-4) & 1.83(-4) &   0.93 & 8.92(-5) &  0.487& xO,xSi\nl
 $^{54}$Cr & 8.08(-6) & 8.06(-6) &   1.00 & 1.11(-10)&  ---  & nse-Ia-MCh\nl
 $^{55}$Mn & 1.12(-3) & 1.11(-3) &   1.00 & 3.39(-4) &  0.305& Ia,xSi,$\nu$\nl
 $^{54}$Fe & 6.94(-3) & 7.31(-3) &   0.95 & 3.51(-3) &  0.480& Ia,xSi\nl
 $^{56}$Fe & 9.65(-2) & 9.61(-2) &   1.00 & 1.30(-1) &  1.353& xSi,Ia\nl
 $^{57}$Fe & 3.45(-3) & 3.44(-3) &   1.00 & 4.64(-3) &  1.349& xSi,Ia\nl
 $^{58}$Fe & 1.58(-4) & 1.60(-4) &   0.99 & 3.82(-10)&  ---  & He(s),nse-Ia-MCh\nl
 $^{59}$Co & 3.00(-4) & 2.81(-4) &   1.07 & 1.36(-4) &  0.484& He(s),$\alpha$,Ia,$\nu$\nl
 $^{58}$Ni & 4.22(-3) & 4.34(-3) &   0.97 & 6.64(-3) &  1.529& $\alpha$\nl
 $^{60}$Ni & 2.50(-3) & 2.67(-3) &   0.94 & 3.13(-3) &  1.172& $\alpha$,He(s)\nl
 $^{61}$Ni & 1.52(-4) & 1.49(-4) &   1.02 & 1.46(-4) &  0.980& $\alpha$,Ia-det,He(s)\nl
 $^{62}$Ni & 7.11(-4) & 8.91(-4) &   0.80 & 1.00(-3) &  1.122& $\alpha$,He(s)\nl
 $^{64}$Ni & 4.36(-5) & 4.59(-5) &   0.95 & 9.17(-16)&  ---  & He(s) \nl
 $^{63}$Cu & 4.01(-5) & 3.16(-5) &   1.27 & 9.56(-7) &  0.030& C,Ne\nl
 $^{65}$Cu & 1.27(-5) & 1.42(-5) &   0.89 & 7.69(-7) &  0.054& He(s)\nl
 $^{64}$Zn & 2.84(-5) & 2.67(-5) &   1.06 & 1.41(-5) &  0.528& $\nu$-wind,$\alpha$,He(s)\nl
 $^{66}$Zn & 5.56(-5) & 5.78(-5) &   0.96 & 1.47(-5) &  0.254& He(s),$\alpha$,nse-Ia-MCh\nl
 $^{67}$Zn & 3.27(-6) & 3.54(-6) &   0.92 & 1.94(-8) &  0.005& He(s)\nl
 $^{68}$Zn & 1.63(-5) & 1.75(-5) &   0.93 & 6.35(-9) &  ---  & He(s)\nl
 $^{70}$Zn & 2.81(-7) & 3.12(-7) &   0.90 & 3.19(-21)&  ---  & He(s)\nl
 $^{69}$Ga & 2.35(-6) & 2.23(-6) &   1.05 & 6.47(-13)&  ---  & He(s)\nl
 $^{71}$Ga & 1.21(-6) & 1.30(-6) &   0.93 & 2.29(-20)&  ---  & He(s)\nl
 $^{70}$Ge & 7.32(-6) & 5.53(-6) &   1.32 & 5.13(-15)&  ---  & He(s)\nl
\tablenotetext{a}{Nucleosynthetic process(es) responsible for the production
 where
``BB'' stands for the big bang (e.g. Walker et al. 1991);
``L*'' stands for low-mass stars (M$\le 8$ M\sun \ e.g. Renzini \& Voli 1981)
or in some cases AGB stars (Sackmann \& Boothroyd 1992);
``$\nu$'' is the neutrino process (e.g. Woosley et al. 1990);
``$\nu$-wind is the neutrino-driven wind from young neutron stars in Type II and Ib
supernovae (e.g. Hoffman, Woosley, \& Qian 1997);
``$\alpha$'' is the $\alpha$-rich freeze out from nuclear statistical equilibrium (e.g.
Woosley, Arnett, \& Clayton 1973);
``Novae'' indicates classical novae (e.g. Woosley 1986);
``Ia'' is Type Ia supernovae (Nomoto, Thielemann, \& Yokoi 1984 for ordinary
Ia's which are assumed to ignite at the Chandrasekhar mass; Woosley \& Eastman 1995
for the inner neutron-rich regions of ordinary Ia's- here called ``nse-Ia-MCh'';
Woosley \& Weaver 1994 for sub-Chandrasekhar detonation models ``Ia-He-det'').
All other entries refer to burning processes in massive stars, H, He, C, Ne, O, Si-
responsible for producing the isotope, with a prefix ``x'' indicating explosive
nucleosynthesis.}
\enddata
\end{deluxetable}

\begin{deluxetable}{l c c c c c l}
\tablewidth{0pt}
\tablecaption{\label{radkepvstnh}Radioactive Stellar Yields (in M\sun)
in the 15 M\sun Supernovae}
\tablehead{
\colhead{$^{\rm A}$Z}&
\colhead{WFHZ}&
\colhead{TAT}&
\colhead{WFHZ/TAT}&
\colhead{TNH}&
\colhead{TNH/TAT}&
\colhead{process(es)$^a$}
}
\startdata
 $^{22}$Na & 8.10(-8) & 9.68(-8) &   0.84 & 3.98(-8) &  0.411& Novae\nl
 $^{26}$Al & 3.07(-5) & 2.92(-5) &   1.05 & 2.68(-6) &  0.092& $\nu$,xNe,H\nl
 $^{44}$Ti & 6.06(-5) & 7.40(-5) &   0.82 & 7.19(-5) &  0.972& $\alpha$\nl
 $^{56}$Ni & 8.11(-2) & 8.06(-2) &   1.01 & 1.30(-1) &  1.613& $\alpha$,nse-Ia-MCh\nl
 $^{60}$Fe & 3.35(-6) & 5.51(-6) &   0.61 & 1.25(-17)&  ---  & Ne,xNe,xHe\nl
 $^{60}$Co & 5.63(-6) & 2.60(-6) &   2.17 & 4.14(-14)&  ---  & Ne,xNe,xHe\nl
\tablenotetext{a}{see footnote table \ref{new15vstnh96}}
\enddata
\end{deluxetable}

\clearpage

\begin{deluxetable}{l c c c c c }
\tablewidth{0pt}
\tablecaption{\label{thvexp}Statistical Errors of Reaction Ratios}
\tablehead{
\colhead{($j,k$)}& 
\colhead{\#}&
\colhead{\=x$_{WFHZ}$}& 
\colhead{$\sigma_{WFHZ}$}&
\colhead{\=x$_{TAT}$}&
\colhead{$\sigma_{TAT}$}
}
\startdata
(n,$\gamma$)& 62& 0.91& 0.58& 1.08& 0.64\nl
(p,n)       & 12& 1.47& 0.62& 0.95& 0.47\nl
(p,$\gamma$)& 20& 1.35& 0.81& 1.33& 0.65\nl
($\alpha$,p)& 12& 0.81& 0.35& 0.73& 0.45\nl
($\alpha$,n)& 16& 0.85& 0.39& 0.86& 0.23\nl
($\alpha$,$\gamma$)& 2& 1.10& 0.36& 1.25& 0.25\nl
\tablenotetext{a}{\=x and $\sigma$ indicate the arithmetic mean and standard deviation
for the stated number (\#) of cross section [(n,$\gamma$)] and reaction
rate (all other channel) ratios (theory/experiment).}
\enddata
\end{deluxetable}

\clearpage

\begin{deluxetable}{l c c c c c }
\tablewidth{0pt}
\tablecaption{\label{vgl}Comparison of stellar reaction rates in cm$^3$ s$^{-1}$
mole$^{-1}$ from the WFHZ, TAT, and NON-SMOKER reaction rate libraries.}
\tablehead{
\colhead{Reaction}&
\colhead{$T_9$}&
\colhead{WFHZ}&
\colhead{TAT}&
\colhead{TAT/WFHZ}&
\colhead{NON-SMOKER}
}
\startdata
$^{32}$S(n,$\gamma$)$^{33}$S$^a$&   3.5& 5.4(5)& 6.9(5)& 1.28& 8.4(5)\nl
$^{33}$S(p,$\gamma$)$^{34}$Cl&      3.5& 5.6(3)& 1.1(4)& 1.96& 6.2(3)\nl
$^{33}$S(n,$\alpha$)$^{30}$Si&      3.5& 5.1(7)& 3.4(7)& 0.67& 3.4(7)\nl
$^{39}$K(n,$\gamma$)$^{40}$K$^a$ &  1.0& 1.4(6)& 1.7(6)& 1.21& 1.8(6)\nl
$^{40}$K(n,$\alpha$)$^{37}$Cl&      1.0& 1.1(7)& 8.5(6)& 0.77& 7.6(6)\nl
$^{40}$K(n,p)$^{40}$Ar       &      1.0& 4.5(6)& 1.8(6)& 0.40& 1.8(6)\nl
$^{45}$Ca(n,$\gamma$)$^{46}$Ca&     1.0& 1.6(6)& 2.5(6)& 1.56& 1.7(6)\nl
$^{46}$Ca(n,$\gamma$)$^{47}$Ca$^a$& 1.0& 7.8(5)& 7.9(5)& 1.01& 4.7(5)\nl
$^{44}$Ti($\alpha$,p)$^{47}$V&      2.5&9.2(-3)&9.6(-3)& 1.04&9.6(-3)\nl
$^{45}$Ti(n,$\alpha$)$^{42}$Ca&     3.5& 6.4(6)& 9.9(6)& 1.55& 1.0(7)\nl
$^{45}$Ti(n,p)$^{45}$Sc       &     3.5& 1.7(8)& 2.4(8)& 1.41& 2.5(8)\nl
$^{43}$Ca(p,$\gamma$)$^{44}$Sc&     3.5& 3.9(4)& 3.5(4)& 0.90& 3.2(4)\nl
$^{44}$Sc(p,$\gamma$)$^{45}$Ti&     3.5& 1.2(4)& 2.5(4)& 2.08& 1.6(4)\nl
$^{50}$V(n,$\gamma$)$^{51}$V&       3.5& 2.8(6)& 2.1(6)& 0.75& 3.9(6)\nl
$^{69}$Zn(n,$\gamma$)$^{70}$Zn&     1.0& 6.6(6)& 1.3(7)& 1.97& 1.1(7)\nl
$^{70}$Zn(n,$\gamma$)$^{71}$Zn&     1.0& 1.4(6)& 2.4(6)& 1.71& 2.6(6)\nl
\tablenotetext{a}{Actual rate used by TAT and WFHZ is based 
on Bao \& K\"appeler 1987}
\enddata
\end{deluxetable}

\clearpage

\begin{deluxetable}{l c c c c c }
\tablewidth{0pt}
\tablecaption{\label{agoscn}Comparison of stellar ($\alpha,\gamma$)
reaction rates
(in cm$^3$ s$^{-1}$ mole$^{-1}$) on self-conjugate nuclei
derived from the CRSEC, SMOKER, and NON-SMOKER statistical model codes.}
\tablehead{
\colhead{Reaction}&
\colhead{$T_9$}&
\colhead{CRSEC}&
\colhead{SMOKER}&
\colhead{SM/CR}&
\colhead{NON-SMOKER}
}
\startdata
$^{24}$Mg($\alpha,\gamma$)$^{28}$Si& 2.5& 1.1(1) & 3.7(1) &  3.36& 2.8(0)\nl
$^{28}$Si($\alpha,\gamma$)$^{32}$S & 2.5& 3.8(-1)& 4.5(0) & 11.71& 4.6(-1)\nl
$^{32}$S($\alpha,\gamma$)$^{36}$Ar & 2.5& 2.3(-1)& 8.9(-1)&  3.87& 9.4(-2)\nl
$^{36}$Ar($\alpha,\gamma$)$^{40}$Ca& 2.5& 3.4(-3)& 2.4(-2)&  7.06& 1.6(-3)\nl
$^{40}$Ca($\alpha,\gamma$)$^{44}$Ti& 2.5& 1.5(-2)& 7.8(-2)&  5.70& 3.3(-3)\nl
$^{44}$Ti($\alpha,\gamma$)$^{48}$Cr& 2.5& 3.3(-6)& 1.6(-4)& 48.48& 1.8(-5)\nl
\enddata
\end{deluxetable}


\begin{thebibliography}{}

\bibitem[Angulo, 1999]{ang99}
Angulo, C., Arnould, M., Rayet, M., (and 25 others) 1999, Nucl. Phys. A, in press.
see also the website http://pntpm.ulb.ac.be/Nacre/nacre.htm

\bibitem[Audi and Wapstra 1995]{audi}
Audi, G. and Wapstra, A.H. 1995, Nucl. Phys. A595, 409

\bibitem[Aufderheide 1991]{aufd91}
Aufderheide, M. B., Baron, E., \& Thielemann, F.-K. 1991, ApJ, 370, 630

\bibitem[Aufderheide et al. 1994]{aufd94}
Aufderheide, M. B., Fushiki, I., Woosley, S. E., \& Weaver, T. A. 1994, 
ApJS, 91, 389

\bibitem[Bao \& K\"appeler 1987]{bk87}
Bao, Z.Y., \& K\"appeler, F. 1987, ADNDT, 36, 411

\bibitem[Beer, Voss, \& Winters 1992]{bwv92}
Beer, H., Voss, F., \& Winters, R. R. 1992, ApJS, 80, 403
 
\bibitem[Berman 1975]{berman75}
Berman, B.L. 1975, At. Data Nucl. Data Tables, 15, 319

\bibitem[Berman \& Fultz 1975]{bf75}
Berman, B.L., \& Fultz, S.C. 1975, Rev. Mod. Phys., 47, 713

\bibitem[Berman et al. 1979]{berman79}
Berman, B.L., Faul, D.D., Alvarez, R.A., Meyer, P., \& Olson, D.L. 1979,
Phys. Rev. C, 19, 1205

\bibitem[Blatt \& Weisskopf 1952]{bw52}
Blatt, J.M., \& Weisskopf, V.F. 1952, {\it Theoretical Nuclear Physics},
(Wiley, New York)

\bibitem[Brown and Rho 1981]{bro81}
Brown, G.E., and Rho, M., 1981, Nuc. Phys., A372, 397

\bibitem[Buchmann 1997]{buch97}
Buchmann, L. 1997, ApJ, 479, 153

\bibitem[Burrows \& Hayes 1996]{bh96}
Burrows, A., \& Hayes, J. 1996, Phys. Rev. Let. 76, 352

\bibitem[Carlos et al. 1974]{cetal74}
Carlos, P., Bergere, R., Heil, H., Lepetre, A., \& Veyssiere, A. 1974,
Nucl. Phys., A219, 61

\bibitem[Catara 1997]{cat97}
Catara, F. et al. 1997, Nuc. Phys., A624, 449

\bibitem[Caughlan \& Fowler 1988]{cf88}
Caughlan, G. A. \& Fowler, W. A. 1998, ADNDT, 40, 283 (CF88)

\bibitem[Caughlan et al. 1985]{cfhz85}
Caughlan, G.R., Fowler, W.A., Harris, M.J., \& Zimmerman, B.A. 1985, ADNDT, 32,
197 

\bibitem[Chieffi, Limongi, \& Straniero 1998]{cls98}
Chieffi, A., Limongi, M., \& Straniero C. 1998, ApJ, 502, 737

\bibitem[Cooperman, Shapiro, \& Winkler 1977]{csw77}
Cooperman, E. L., Shapiro, M. H., \& Winkler, H. 1977, Nucl. Phys. A284, 163

\bibitem[Cowan, Thielemann, \& Truran 1991]{ctt91}
Cowan, J.J., Thielemann, F.-K., Truran, J. W. 1991, Phys. Rep., 208, 267

\bibitem[Dean et al.\ 1998]{dean98}
Dean, D.J., Langanke, K., Chatterjee, L., Radha, P.B., and Strayer,
M.R., 1998, Phys.\ Rev.\ C 58, 536

\bibitem[Dixon \& Storey, 1977]{ds77}
Dixon, W. R. \& Storey, R. S. 1977, Phys. Rev. C, 15, 1896 

\bibitem[Firestone 1996]{ENSDF}
Firestone, R.B., 1996, {\it Table of Isotopes, 8th Edition}, John Wiley
\& Sons, New York

\bibitem[Ezhov \& Plukjo 1993]{ep93}
Ezhov, S.N., \& Plujko, V.A. 1993, Z. Phys., 346, 275

\bibitem[Fantoni, Friman, \& Pandharipande 1981]{fetal81}
Fantoni, S., Friman, B.L., \& Pandharipande, V.R. 1981, Phys. Rev.
Lett., 1048, 89

\bibitem[Fuller, Fowler, \& Newman 1980]{ffn80}
Fuller, G.M., Fowler, W.A., \& Newman, M.J. 1980, ApJS, 42, 447

\bibitem[Fuller, Fowler, \& Newman 1982a]{ffn82a}
Fuller, G.M., Fowler, W.A., \& Newman, M.J. 1982, ApJS, 42, 447

\bibitem[Fuller, Fowler, \& Newman 1982b]{ffn82b}
Fuller, G.M., Fowler, W.A., \& Newman, M.J. 1982, ApJ, 252, 715

\bibitem[Fuller, Fowler, \& Newman 1985]{ffn85}
Fuller, G.M., Fowler, W.A., \& Newman, M.J. 1985, ApJ, 293, 1

\bibitem[Fowler, Caughlan, \& Zimmerman 1975]{fcz75}
Fowler, W.A., Caughlan, G.R., \& Zimmerman, B.A. 1975, ARAA, 13, 69

\bibitem[Gadioli \& Hodgson 1992]{gh92}
Gadioli, E., \& Hodgson, P.E., 1992, {\it Pre-Equilibrium
Nuclear Reactions }, Clarendon Press, Oxford

\bibitem[Gilbert \& Cameron 1965]{gc65}
Gilbert, A., \& Cameron, A.G.W. 1965, Can. J. Phys., 43, 1446

\bibitem[Gurevich et al. 1981]{gur81}
Gurevich, G.M. et al. 1981, Nucl. Phys., A351, 257

\bibitem[Hardy 1982]{hardy82}
Hardy, J.C. 1982, Phys. Lett., 109B, 242

\bibitem[Harris et al. 1983]{hfcz83}
Harris, M.J., Fowler, W.A., Caughlan, G.R., \& Zimmerman, B.A. 1983,
ARAA, 21, 165

\bibitem[Hashimoto, Hanawa, \& Sugimoto 1983]{hhs83}
Hashimoto, K., Hanawa, T., \& Sugimoto, D. 1983, Publ. Astron. Soc. Japan 35, 1

\bibitem[Hashimoto et al 1993]{hash93}
Hashimoto, M., Iwamoto, K., \& Nomoto, K. 1993, ApJ, 414, 105

\bibitem[Hashimoto 1985]{hash95}
Hashimoto, M. 1995, Prog. Theor. Phys., 94, 663

\bibitem[Hix \& Thielemann 1996]{hixt96}
Hix, R. \& Thielemann, F.-K. 1996, ApJ, 460, 869

\bibitem[Hix \& Thielemann 1996]{hixt98}
Hix, R. \& Thielemann, F.-K. 1998, ApJ, in press

\bibitem[Hoffman \& Woosley 1992]{hw92}
Hoffman, R. D. \& Woosley S. E. 1992, 
``Tables of Thermonuclear Reaction Rates for Nucleosynthesis (Z $<$ 44)",
V. 92.1, unpublished, see http://ie.lbl.gov/astro/astrorate.html

\bibitem[Hoffman, Woosley, \& Qian 1997]{hwq97}
Hoffman, R. D., Woosley S. E., \& Qian, Y.-Z. 1997, ApJ, 482, 951

\bibitem[Holmes 1976]{holmes76}
Holmes, J. A. 1976, PhD thesis, Cal-Tech, unpublished

\bibitem[Holmes et al. 1976]{hetal76}
Holmes, J. A., Woosley, S.E., Fowler, W.A., \& Zimmerman, B.A. 1976,
ADNDT, 18, 305

\bibitem[Janka \& M\"uller 1996]{jm96}
Janka, H.-T., \& M\"uller, E. 1996, A\&A, 306, 167

\bibitem[Jeukenne, Lejeune, \& Mahaux 1977]{jlm77}
Jeukenne, J.P., Lejeune, A., \& Mahaux, C. 1977, Phys. Rev. C, 16, 80

\bibitem[Johnson 1977]{johnson77}
Johnson, C.H. 1977, Phys. Rev. C, 16, 2238

\bibitem[K\"appeler et al. 1994]{kaplr94}
K\"appeler, F., Wiescher, M., Giesen, U., Goerres, J., Baraffe, I.,
El Eid, M., Raiteri, C. M., Busso, M., Gallino, R., \& Chieffi, A. 1994,
ApJ, 437, 396

\bibitem[Mahaux \& Weidenm\"uller 1979]{m79}
Mahaux, C., \& Weidenm\"uller, H.A. 1979, Ann. Rev. Part. Nucl. Sci., 29, 1

\bibitem[Mahaux 1982]{mah82}
Mahaux, C. 1982, Phys. Rev. C, 82, 1848

\bibitem[Malaney \& Fowler 1988]{mf88}
Malaney, R. A. \& Fowler, W. A. 1988, ApJ, 333, 14

\bibitem[Malaney \& Fowler 1989]{mf89}
Malaney, R. A. \& Fowler, W. A. 1989, ApJ, 345, 5

\bibitem[Mann, 1978]{mann78}
Mann, F.M. 1978, Hauser 5, A Computer Code to Calculate Nuclear Cross
Sections, Hanford Engineering (HEDL-TME 78-83)

\bibitem[Mezzacappa et al. 1997]{metal97}
Mezzacappa, A., Calder, A.C., Bruenn, S.W., Blondin, J.M., Guidry, M.W.,
   Strayer, M.R., Umar, A.S., 1997, Ap. J., in press

\bibitem[McFadden \& Satchler 1966]{ms66}
McFadden, L., \& Satchler, G.R. 1966, Nucl. Phys., 84, 177

\bibitem[Michaud, Scherk, \& Vogt 1970]{mic70}
Michaud, G., Scherk, L., \& Vogt, E. 1970, Phys. Rev. C, 1, 864

\bibitem[M\"oller et al. 1995]{moeller95}
M\"oller, P., Nix, J.R., Myers, W.D., \& Swiatecki, W.J. 1995,
ADNDT, 59, 185

\bibitem[M\"oller, Nix, \& Kratz 1997]{moellkra}
M\"oller, P., Nix, J.R., \& Kratz, K.-L. 1997,
ADNDT, 66, 131

\bibitem[Myers et al. 1977]{myers77}
Myers, W.D., Swiatecki, W.J., Kodama, T., El-Jaick, L.J., Hilf, E.R. 1977,
Phys. Rev. C, 15, 2032

\bibitem[Nakada and Alhassid 1997]{na97}
Nakada, H., \& Alhassid, Y. 1997, Phys. Rev. Lett., 79, 2939

\bibitem[Nomoto, Thielemann, \& Yokoi 1984]{nty84}
Nomoto, K., Thielemann, F.-K., \& Yokoi, Y. 1984, A\&A, 286, 644

\bibitem[Nomoto, \& Hashimoto 1988]{nh88}
Nomoto, K., \& Hashimoto, M. 1988, Phys. Rep., 163, 13

\bibitem[Nomoto et al. 1997]{netal97}
Nomoto, K., Hashimoto, M., Tsujimoto, T., Thielemann, F.-K., Kishimoto,
N., Kubo, Y., Nakasato, N. 1997, Nucl. Phys., A161, 79c

\bibitem[Peaslee 1957]{pea57}
Peaslee, D.C. 1957, Nucl. Phys. 3, 255

\bibitem[Prantzos, Hashimoto, \& Nomoto 1990]{phn90}
Prantzos, N., Hashimoto, M., \& Nomoto, K. 1990, A\&A, 234, 211

\bibitem[Prantzos et al. 1993]{pran93}
Prantzos, N., Casse, M., \& Vangioni-Flam, E. 1993, ApJ, 403, 630

\bibitem[Rogers, Dixon, \& Storey, 1977]{rds77}
Rogers, D. W. O., Dixon, W. R., \& Storey, R. S. 1977, Nucl. Phys. A281, 345

\bibitem[Rauscher, Thielemann \& Kratz 1997]{rtk97}
Rauscher, T., Thielemann, F.-K., \& Kratz, K.-L. 1997, Phys. Rev. C, 56, 1613

\bibitem[Rauscher, Thielemann, \& Kratz 1997]{rau97}
Rauscher, T., and Thielemann, F.-K., \& K.-L. Kratz, 1997, Phys.
Rev. C, 56, 1613

\bibitem[Rauscher 1998]{rau98}
Rauscher, T. 1998, in {\it ``Nuclear Astrophysics''}, eds.\ M. Buballa,
W. N\"orenberg, J. Wambach, A. Wirzba (Gesellschaft f\"ur
Schwerionenforschung (GSI), Darmstadt 1998), p.\ 288; nucl-th/9802026

\bibitem[Rauscher and Thielemann 1998]{rauoak}
Rauscher, T., and Thielemann, F.-K. 1998, in {\it Stellar Evolution,
Stellar Explosions, and Galactic Chemical Evolution}, ed.\ A. Mezzacappa
(IOP, Bristol 1998), p.\ 519;
see also http://quasar.physik.unibas.ch/\~{ }tommy/reaclib.html\#refs

\bibitem[Renzini \& Voli 1981]{rv81}
Renzini, A., \& Voli, M. 1981, A\&A, 94, 175

\bibitem[Rayet et al. 1995]{rayet95}
Rayet, M., Arnould, M., Hashimoto, M., Prantzos, N., \& Nomoto, K. 1995,
A\&A, 298, 517

\bibitem[Sachmann \& Boothroyd 1992]{sb92}
Sachmann, I.J., \& Boothroyd, A.I. 1992, ApJ, L71

\bibitem[Sargood 1982]{sar82}
Sargood, D. G. 1982, Phys. Rep. 93, \#2, 61

\bibitem[Satchler \& Love 1979]{sat79}
Satchler, G.R., \& Love, W.G. 1979, Phys. Rep., 55, 183

\bibitem[Somorjai 1998]{sam98}
Somorjai, F. et al. 1998, A\&A 333, 1112

\bibitem[Tepel, Hoffmann, \& Weidenm\"uller 1974]{tepel74}
Tepel, J.W., Hoffmann, H.M., \& Weidenm\"uller, H.A. 1974, Phys. Lett., 49B, 1

\bibitem[The et al. 1998]{tcjm98}
The, L.-S., Clayton, D. D., Jin, L., \& Meyer, B. S. 1998, ApJ, 504, 500 

\bibitem[Thielemann \& Arnould 1983]{ta83}
Thielemann, F.-K, \& Arnould, M. 1983, in {\it Proc. Int. Conf. on Nucl.
Data for Sci. and Technology}, ed. K. B\"ockhoff (Reidel, Dordrecht), p.762

\bibitem[Thielemann \& Arnett 1985]{ta85}
Thielemann, F.-K., \& Arnett, D. 1985, ApJ, 295, 604

\bibitem[Thielemann, Arnould, \& Truran 1987]{tat87}
Thielemann, F.-K., Arnould, M., \& Truran, J. 1987, in {\it Advances in
Nuclear Astrophysics}, ed.\ E. Vangioni-Flam (Editions Fronti\`ere,
Gif sur Yvette 1987), p.\ 525 (TAT)

\bibitem[Thielemann, Hashimoto \& Nomoto 1990]{thn90}
Thielemann, F.-K., Hashimoto, M., \& Nomoto, K. 1990, ApJ, 349, 222 

\bibitem[Thielemann, Nomoto, \& Hashimoto 1996]{tnh96}
Thielemann, F.-K., Nomoto, K., \& Hashimoto, M. 1996, ApJ, 460, 408 (TNH)

\bibitem[Thomas, Zirnbauer \& Langanke 1986]{tzl86}
Thomas, J., Zirnbauer, M.R., \& Langanke, K. 1986, Phys. Rev. C, 33, 2197

\bibitem[Timmes, Woosley, \& Weaver 1995]{tww95}
Timmes, F. X., Woosley, S. E., \& Weaver, T. A. 1995, ApJS, 98, 617

\bibitem[Toevs, 1977]{toevs71}
Toevs, J. W. 1971, Nucl. Phys., A172, 589 

\bibitem[Tuli et al. 1990]{nwc90}
J. Tuli, 1990, Nuclear Wallet Cards, (National Nuclear Data
Center, Brookhaven National Lab.)

\bibitem[Van Wormer 1994]{vanw94}
Van Wormer, L., Goerres, J., Iliadis, C., Wiescher, M., \& Thielemann, F.-K.
1994, ApJ, 432, 326

\bibitem[Verbaatschot et al.\ 1984]{ver84}
Verbaatschot, J.J.M., Weidenm\"uller, H.A., and Zirnbauer, M.R., 1984,
Phys.\ Rep.\ 129, 367

\bibitem[Vogt 1968]{vog68}
Vogt, E. 1968, in {\it Advances in Nuclear Physics, Vol. 1},
eds.\ M.Baranger, E. Vogt (Plenum Press, New York), p. 261

\bibitem[Walker et al. 1991]{walker91}
Walker, T.P., Steigman, G., Schramm, D.N., Olive, K.A., \& Kang, H. 1991,
ApJ, 376, 51

\bibitem[Weaver, Zimmerman, \& Woosley 1978]{wzw78}
Weaver, T.A., Zimmerman, G.B., \& Woosley, S.E. 1978, ApJ, 225, 1021

\bibitem[Weaver \& Woosley 1993]{ww93}
Weaver, T.A., \& Woosley, S.E. 1993, Phys. Rep., 227, 65

\bibitem[Wiescher et al. 1986]{wies86}
Wiescher, M., Goerres, J., Thielemann, F.-K., \& Ritter, H. 1986, A\&A, 160, 56

\bibitem[Wiescher 1999]{wies99}
Wiescher, M. 1999, private communication.

\bibitem[Wiescher et al. 1987]{wies87}
Wiescher, M., Harms, V., Goerres, J., Thielemann, F.-K., \& 
Rybarcyk, L. J. 1987, ApJ, 316, 162

\bibitem[Wiescher et al. 1989]{wies89}
Wiescher, M., Goerres, J., Graff, S., Buchmann, L., \& Thielemann, F.-K. 1989,
ApJ, 343, 352

\bibitem[Wisshak et al. 1997]{wetal97}
Wisshak, K., Voss, F., K\"appeler, F., and Kerzakov, I., 1997, Nucl.\
Phys. A621, 270c

\bibitem[Woosley, Arnett, \& Clayton 1973]{wac73}
Woosley, S.E., Arnett, D., \& Clayton, D. D. 1973, ApJS, 26, 231

\bibitem[Woosley et al. 1976]{opa422}
Woosley, S.E., Fowler, W.A., Holmes, J.A., \& Zimmerman, B.A. 1975,
Tables of Thermonuclear Reaction Rate Data for Intermediate
Mass Nuclei, (OAP-422), unpublished

\bibitem[Woosley et al. 1978]{wfhz78}
Woosley, S.E., Fowler, W.A., Holmes, J.A., \& Zimmerman, B.A. 1978,
ADNDT, 22, 371 (WFHZ)

\bibitem[Woosley \& Weaver 1978]{ww78}
Woosley, S.E. \& Weaver, T.A. 1978, ApJS, 101, 181

\bibitem[Woosley 1986]{woos86}
Woosley, S.E. 1986, in Nucleosynthesis and Chemical Evolution, 16th 
Advanced Course Swiss Society of Astrophysics and Astronomy, ed. B.
Hauck, A. Maeder, \& G. Meynet (Geneva: Geneva Obs.), 74

\bibitem[Woosley \& Weaver 1986]{ww86}
Woosley, S.E. \& Weaver, T.A. 1986, ARA\&A, 24, 205

\bibitem[Woosley et al. 1990]{whhh90}
Woosley, S. E., Hartmann, D. H., Hoffman, R. D., \& Haxton, W. C. 1990, 
ApJ, 356, 272

\bibitem[Woosley & Hoffman 1992]{wh92}
Woosley, S. E. \& Hoffman, R. D. 1992, ApJ, 395, 202

\bibitem[Woosley \& Eastman 1995]{we95}
Woosley, S. E. \& Eastman, R. E. 1995, in Type I Supernovae,
Proc. 1993, Menorca Summer School, ed. E. Bravo, I. Ibanez, \& J. Isern (Singapore:
World Scientific), 220

\bibitem[Woosley \& Weaver 1994]{ww94}
Woosley, S.E. \& Weaver, T.A. 1994, ApJ, 423, 371 

\bibitem[Woosley \& Weaver 1995]{ww95}
Woosley, S.E. \& Weaver, T.A. 1995, ApJS, 101, 181 (WW95)

\bibitem[Woosley \& Weaver 1999]{ww99}
 Woosley, S. E., \& Weaver, T. A. 1999, Rev Mod Phys, in preparation

\end{thebibliography}
\end{document}